\documentclass{aastex}
\usepackage{emulateapj5}
\usepackage{onecolfloat}
\usepackage{epsf}

\newcommand{\kpe}{k_\perp}
\newcommand{\kpa}{k_\parallel}
\newcommand{\Om}{\Omega_m}
\newcommand{\LCDM}{\rm {\Lambda CDM}}
\newcommand{\OL}{\Omega_\Lambda}
\newcommand{\Ok}{\Omega_K}
\newcommand{\DA}{D\!_A(z)}
\newcommand{\hz}{H(z)}
\newcommand{\DAA}{D_A\,}
\newcommand{\hzz}{H\,}

\newcommand{\Vsur}{V_{\rm survey}}
\newcommand{\Veff}{V_{\rm eff}}
\newcommand{\Oh}{\Omega_m h^2}
\newcommand{\Ob}{\Omega_b h^2}
\newcommand{\w}{w}
\newcommand{\wpivot}{w(z_{\rm pivot})}
\newcommand{\Ox}{\Omega_X}
\newcommand{\hMpc}{h^{-1}{\rm\;Mpc}}
\newcommand{\ihMpc}{h{\rm\;Mpc^{-1}}}
\newcommand{\kmax}{k_{\rm max}}
\newcommand{\kfid}{k_{\rm fid}}
\newcommand{\kn}{k_{\rm norm}}
\newcommand{\Mpc}{{\rm\;Mpc}}
\newcommand{\Gpc}{{\rm\;Gpc}}
\newcommand{\iMpc}{{\rm\;Mpc^{-1}}}
\newcommand{\DAcmb}{D_{A,\rm CMB}}

\newcommand{\physlet}{Phys Lett}

\newcommand{\tableskni}{&&&&&&&&\\[-7pt]}
\newcommand{\tableskte}{&&&&&&&&&\\[-10pt]}
\newcommand{\tableskten}{&&&&&&&&&\\[-7pt]}
\newcommand{\tableskth}{&&\\[-7pt]}
\newcommand{\tableskei}{&&&&&&& \\[-7pt]}
\newcommand{\tableskft}{&&&&&&&&&&&&&&\\ [-7pt]}
\begin{document}
\twocolumn[%
\submitted{Submitted to \textit{The Astrophysical Journal} 5-26-2003} 
\title{Probing Dark Energy with Baryonic Acoustic Oscillations \\ from Future Large Galaxy Redshift Surveys} 
\author{Hee-Jong Seo \& Daniel J. Eisenstein \\ \protect }

\begin{abstract}
We show that the measurement of the baryonic acoustic oscillations
in large high redshift galaxy surveys offers a precision route to the
measurement of dark energy. The cosmic microwave background provides the
scale of the oscillations as a standard ruler that can be measured in
the clustering of galaxies, thereby yielding the Hubble parameter and
angular diameter distance as a function of redshift. This, in turn,
enables one to probe dark energy. We use a Fisher matrix formalism
to study the statistical errors for redshift surveys up to $z=3$ and
report errors on cosmography while marginalizing over a large number of
cosmological parameters including a time-dependent equation of state. With
redshifts surveys combined with cosmic microwave background satellite
data, we achieve errors of 0.037  on $\Ox$, 0.10 on $\w(z=0.8)$,
and 0.28 on $d\w(z)/dz$ for cosmological constant model. Models with
less negative $w(z)$ permit tighter constraints. We test and discuss the
dependence of performance on redshift, survey conditions, and fiducial
model.  We find results that are competitive with the performance of 
future supernovae Ia
surveys. We conclude that redshift surveys offer a promising independent
route to the measurement of dark energy.
\end{abstract}

]

\section{Introduction}
Recent observations of distant type Ia supernovae have reached the startling conclusion that the expansion of the Universe is accelerating
\citep{Perlm99,Riess98,Rie01,Ton03}. 
Under the premise of Friedmann equations, 
this implies the existence of an energy component, 
christened dark energy, with negative pressure \citep{Rat88,Fri95}. 
The detailed characterization of the accelerated expansion and 
its cause is now one of the main subjects of cosmology. 
Dark energy presently constitutes about $2/3$ of the total energy density of the Universe and its physical property is often parameterized by the ratio of pressure to density, that is, the equation of state
\citep{Ste97,TW97}. 
A cosmological constant \citep[for a review, see][]{Ca92}
has a constant equation of state of $-1$, 
while general quintessence models \citep{Cal98}
and other theories 
\citep{Zla99,Buc99,Arm00,Boy01,Gu01,Kas01,Bil02,Def02,Fre02}
typically allow equations of state 
with a redshift dependence.
Measuring the time dependence of the equation of state, 
as well as its present density, is an essential step in identifying 
the physical origin of dark energy 
\citep{Hui99,Coo99,Hut99,New00,Hai01,Hut01,Maor01,Wan01,Kuj02,Maor02,New02,Wel02,Fri03,Lin03}. 
Because of the inertness and the relatively smoothness of this 
energy component, as commonly believed in the standard pictures 
of dark energy, the best cosmological probe of dark energy is the 
expansion history of the Universe, 
measured by the Hubble parameter and angular diameter distance.

In this paper, we demonstrate that the Hubble parameter 
$\hz$ and angular diameter distance $\DA$ can be measured to 
excellent precision by using the baryonic acoustic oscillations 
imprinted in the large-scale structure of galaxies. 
We are familiar with this signature as the now-famous Doppler peaks 
in the anisotropies of the cosmic microwave background
\citep{Peebles70,Bon84,Mil99,deB00,Han00,HalDasi,BenoitArcheops,BennettWmap}; 
however, the same structure is predicted to be present in 
the late-time clustering of galaxies
as a series of weak modulations in the amplitude of fluctuations 
as a function of scale
\citep{Peebles70,Bon84,Holtzman89,HS96}. 
The physical scale of the oscillations is determined by the matter 
and baryon densities, which can be precisely measured with CMB anisotropy data. 
This calibrates the acoustic oscillations as a standard ruler 
\citep{Eht98,Eis03}.
The observed length scales of oscillations in the transverse and 
line of sight directions in a galaxy redshift survey then determine 
the angular diameter distance $\DA$ and the Hubble parameter $\hz$ as 
functions of redshift. As an oscillatory feature, the acoustic signature 
is less susceptible to general systematic errors and distortions; 
however, only large surveys map enough cosmic volume to achieve the 
precision required to detect these features. 
In addition, the features along the line-of-sight clustering are 
on sufficiently small scales that resolving them requires an 
accurate measurement of redshift, motivating the need for
spectroscopic redshift surveys. Surveys at higher redshift are 
preferred so as to avoid the erasure of the oscillatory features 
by nonlinear structure formation \citep{Jai94,Meiksin99,Meiksin99b}.
Recent analyses of large surveys may be beginning to reveal
these features \citep{Per01,Mil01}

There have been numerous studies on how the combination
of CMB anisotropy data and large-scale structure data, either
present \citep{Scott95,Gawiser98,Lan00,Teg01,Efs02,Spe03} 
or future \citep{Hu98b,Eht98,Wang99,Eht99,Pop01}, 
can constrain cosmological parameters.
These studies have considered an increasing number of parameters 
and degeneracies and build on a body of work in CMB parameter
estimation \citep{Kno95,Jun96,Zal97,Bon97}.  
However, most previous work on galaxy surveys has concentrated on low redshifts
and used spherically averaged power spectra. 
The spherical assumption neglects the effects of redshift distortions and 
cosmological distortions.  
Including the non-isotropic information in the clustering of galaxies
allows one to recover these effects \citep{Bal95,Heavens97,Hatton99,Taylor01,Mat02,Mats03}.

In this paper, we design large galaxy redshift surveys at high redshift
that can recover the acoustic peaks with a level of precision that allows us to put competitive constraints on the dark energy. 
We describe the constraints in terms of statistical errors using a Fisher matrix treatment of the full three-dimensional power spectra. 
We study galaxy survey at $z=0.3$, $z\sim 1$, and $z=3$ so as to have access to cosmological distortions across a wide range of cosmic history. 
As our goal is to optimize survey design based on realistic statistical
errors, we try to be conservative in our methodology.
For example, we adopt ungenerous values for the non-linear scales 
and marginalize over a large number of cosmological parameters. 
We present the predicted performance of the our baseline surveys
with constraints derived for $\hz$ and $\DA$ and then propagate 
these errors to the constraints on the dark energy parameters at 
our fiducial cosmology model, $\LCDM$.
This work extends that of \citet{Bla03} in that we have used a 
full Fisher matrix formalism to treat the cosmological constraints
from large-scale structure, CMB
anisotropies, and supernova data simultaneously and that we have
considered time-variable equations of state.  It differs from
\citet{Lin03b} in that it is an explicit treatment of the survey
data sets in addition to a discussion of dark energy parameter 
estimation. Contemporaneously with this paper, \citet{HuH03} used a Fisher matrix technique similar to ours to study the performance of a mid-redshift cluster survey. The two analyses differ in numerous details.

In \S~\ref{sec:physics}, we discuss the details of the physics to probe
dark energy. In \S~\ref{sec:Metho}, we present the survey condition we
assume, and our Fisher information matrix methodology. We present and
discuss our results in \S~\ref{sec:RD}. We consider variations in survey
design, and fiducial model. We compare the performance to a supernovae
survey (SNe) and to pure imaging surveys.

\section{From Baryonic Oscillations to Dark Energy} \label{sec:physics}
\subsection{Cosmography and Dark Energy}\label{subsec:cosmography}

The expansion history of the universe can be written as the redshift $z(t)$
as a function of time, which in turn is completely specified by the 
Hubble parameter $H(z)$ as a function of redshift.  We will probe the
expansion history by measuring $H(z)$ and the angular diameter distance
$\DA$.

The evolution of dark energy density can be described by the 
present-day dark energy density $\Ox$ and the equation of state 
of dark energy, $\w_X(z)$ \citep{Ste97,TW97}, where

\begin{equation} \label{eq:wz0}
\w_X(z)=\left. \frac{p_X}{\rho_X} \right |_z 
\end{equation}

This yields an energy density as a function of redshift

\begin{equation}\label{eq:rhox}
\rho_X(z)=\rho_X(0) \exp \left[ 3 \int_0^z \frac{1+\w(z)}{1+z}dz \right]
\end{equation}

Assuming a flat Universe, $\DA$ and $\hz$ are then related to the dark energy density through
\begin{eqnarray*}
\hz=h \sqrt{\Om(1+z)^3+\Ox \exp \left[3 \int_0^z \frac{1+\w(z)}{1+z}dz\right]} \nonumber 
\end{eqnarray*}
\begin{equation}\label{eq:hzda}
=\sqrt{\frac{\Oh}{1-\Ox}}\sqrt{\Om(1+z)^3+\Ox \exp \left[ 3 \int_0^z \frac{1+\w(z)}{1+z}dz \right]} 
 \end{equation}

\begin{equation}
\DA=\frac{c}{1+z}\int_0^z \frac{dz}{\hz} \label{eq:hzda1}
\end{equation}
where $\Ox$ is the present-day dark energy fraction with respect to the critical density.
In a general sense, $\hz$ and $\DA$ are the fundamental observables, to be interpreted here as $\Ox$ and $\w(z)$. 
 The comoving sizes of an object or a feature at redshift $z$ in line-of-sight ($r_\parallel$) and transverse ($r_\perp$) directions are related to the 
observed sizes $\Delta z$ and $\Delta \theta$ by $\hz$ and $\DA$:  
\begin{eqnarray}
r_\parallel&=&\frac{c\Delta z}{\hz} \label{eq:rp}\\
r_\perp&=& (1+z)\DA \Delta \theta  \label{eq:rp1}
\end{eqnarray}
When the true scales, $r_\parallel$ and $r_\perp$, are known,
measurements of the observed dimensions, $\Delta z$ and $\Delta \theta$,
give estimates of $H(z)$ and $\DA$.  The object is then known as a
``standard ruler.'' Equations (\ref{eq:rp}) and (\ref{eq:rp1}) can be
applied equally well in Fourier space (inverted, of course).

It is well-known that even if we do not know the scale of a feature,
we can still extract the product $H(z)D_A(z)$ \citep{AP79}.
The acoustic oscillation method presented here is not an Alcock-Paczynski
method because we do know the scale of the sound horizon.

The cosmological feature to be measured need not be an actual object. 
Instead, we can use a statistical property of structure 
in many realizations such as correlation length \citep{Bal95,Mats03}. 
On large scales, features in the power spectrum may be more prominent
and hence easier to use.

\begin{figure}[t]
\plotone{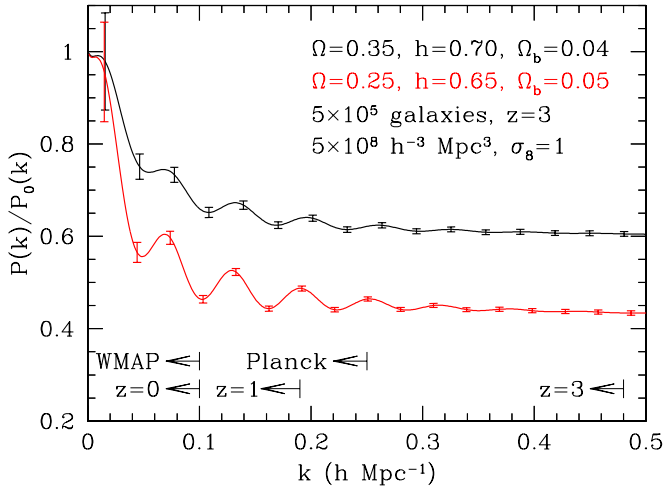}
\caption{The linear power spectrum in two different cosmological models,
$\Omega_m=0.35$, $h=0.70$, and $\Omega_b=0.04$ and
$\Omega_m=0.25$, $h=0.65$, and $\Omega_b=0.05$.
Each power spectrum has been divided by the zero-baryon power
spectrum for that $\Omega_m$ and $h$.
The series of acoustic oscillations is clearly seen.
Lines at the bottom show the non-linear scale, shortward
of which the acoustic oscillations are washed out, as
a function of redshift.  The scales probed by the WMAP
and Planck satellite measurements of primordial anisotropy
are also shown.  The error bars show the spherically averaged
bandpower measurements from the $z=3$ survey we will present
in \S \protect\ref{subsec:sigmap}.
}
\label{fig:bumps}
\end{figure}
 
$\w(z)$ can be written as a derivative of the $\hz$ versus redshift,
which in turn is a derivative of the angular diameter distance versus
redshift (eq.\ [\ref{eq:hzda}] $\&$ eq.\ [\ref{eq:hzda1}]). If we seek not only to measure the
mean value of $\w(z)$ but its slope in redshift, we are adding yet another
derivative to the process. In short, to measure the time variation of
the equation of state, we must be able to measure the second derivative
of $\hz$ or the third derivative of the distance-redshift relation. As
each derivative magnifies the measurement noise in its parent function,
we require enormous precision to proceed.  In the context of galaxy
surveys, this will drive us to require large volumes.

\subsection{Baryonic Acoustic Oscillations in the Matter Power Spectrum}\label{subsec:baryonic}

Baryonic acoustic oscillations are a generic feature of the power
spectrum of large-scale structure and an excellent candidate for the 
standard ruler test.
Prior to recombination, the baryons in the universe are locked to photons 
of the cosmic microwave background, and the photon pressure interacting against 
the gravitational instability produces a series of sound waves in the plasma. 
After recombination, the baryons and photons separate, but the effects of 
the acoustic oscillations remain imprinted in their spatial structure
of the baryons and eventually the dark matter
\citep{Peebles70,Holtzman89,HS96,Eis98a}. 
The resulting power spectrum is shown in Figure \ref{fig:bumps}.

The physical length scale of the acoustic oscillations depends on the sound 
horizon of the universe at the epoch of recombination. 
The sound horizon is the comoving distance a sound wave can travel before recombination and depends simply on the baryon and matter densities. 
The relative heights of the acoustic peaks in the CMB anisotropy power 
spectra measure these densities to excellent accuracy, 
thereby producing an accurate measurement of the sound horizon
\citep{Eht98,Eis03}. 

While the matter power spectrum is simply a product of the spectrum of
primordial fluctuations and the modification of those fluctuations
in later epochs, notably the radiation-domination era and 
recombination, our observations of this power spectrum are 
complicated by the biases of galaxy clustering, the distortions
from peculiar velocities, and the errors induced from reconstructing
distances with the wrong cosmology.  The latter two effects break
the intrinsic statistical isotropy of the clustering of matter
and introduce variations that depend on the angle of the wavevector to
the line of sight.

In the absence of massive neutrinos \citep{Bon83}, linear perturbation
theory fixes the shape of the matter power spectrum in comoving coordinates and changes only the amplitude as the structure evolves.
The growth function $G(z)$ rescales the amplitude of the fixed
matter power spectrum to account for the growth of structure from the
recombination to a redshift $z$.  The growth function does depend on the
details of dark energy.  However, the subtle changes in the amplitude of
the matter power spectrum are easily confused in galaxy redshift surveys
with evolution in the bias of galaxies.  While bias can be estimated from
redshift distortions, recovering it to the $1\%$ accuracy required for
interesting constraints on dark energy is unlikely, especially in light
of the systematic uncertainties of poorly known scale-dependencies of
the redshift distortion.

In principle, galaxy clustering bias could be arbitrary \citep{Dek99}; 
however, under the assumptions of local bias and Gaussian statistics for
the density field, the bias on large scales should be independent
of scale in the correlation function \citep{Coles93,Sch98,Meiksin99,Col99}.
In the power spectrum, this appears as a constant multiplicative bias
plus a constant additive offset \citep{Sel00}.
Moreover, even if the bias deviates from scale independence on 
linear or quasi-linear scales, it is very implausible for it to
introduce oscillations in Fourier space on the acoustic scales,
as this would correspond to a preferred length scale in real space of
enormous size ($\gtrsim30\Mpc$).

Redshift distortions are an angle-dependent distortion in power caused
by the peculiar velocities of galaxies 
\citep[][and references therein]{Hamilton97}.  On the largest scales,
these distortions follow a simple form \citep{Kaiser87} in which
the distortion is an angle-dependent, multiplicative change in power.  
We will follow 
this prescription.  In reality, redshift distortions are non-linear,
including the finger of God effects on small scales.  However, these
deviations have no large preferred length scale and will not
disturb analysis of the acoustic oscillations.

Whereas the linear-theory redshift distortions are an angle-dependent
modulation in the power spectrum amplitude, the cosmological distortion
resulting from an incorrect mapping of observed separations to true
separations produces a distortion in scale.  Spherical features in power
become ellipsoids under the false cosmology.  Were the power spectrum
a simple power law, the cosmological and redshift 
distortions would be indistinguishable
in their quadrupole signatures and difficult to separate overall.
Fortunately, the matter power spectrum is not a simple power law and 
the slow rollover in the power lifts some of the degeneracy between
the two distortions \citep{Bal95,Mats03}.
However, strong features such as baryonic acoustic oscillations are
far more powerful at separating the two, because with a rapidly varying
function, the difference between dilating the scale and modulating
the amplitude is very stark.

Unfortunately, the use of baryonic oscillations as a standard ruler
to derive $\DA$ and $\hz$ is not always straightforward. The nonlinear
gravitational growth of perturbation in the large scale structure erases
the primordial features on smaller scales (large wavenumbers). This occurs
when perturbations on a given scale become of order unity in amplitude,
leading to non-linear coupling between Fourier modes. The obscuration
by nonlinearity moves to a larger scale as the Universe evolves,
and today, the scale corresponds to wavelengths of about 60$\hMpc$,
enough to wipe out all but the first and a part of the second of the
acoustic oscillations \citep{Meiksin99}.  At higher redshift, the
process is less advanced, and we can recover the primordial signals on
smaller scales, including the full series of acoustic oscillations. For
example, at $z=3$, we should be able to recover primordial information
to roughly 12$\hMpc$ (a factor of two smaller than what can be found
in the primary anisotropies of the microwave background), which means that 
many acoustic oscillations can be preserved outside of nonlinearity
region. In practice, we will be limited to about four peaks because 
Silk damping makes the higher harmonics smaller than our expected 
power spectrum measurements.  Figure \ref{fig:bumps} shows the non-linear
scale as a function of redshift, as well as the scales probed by the
CMB primary anisotropies as measured by the WMAP and Planck satellites.
While low-redshift surveys such as the Sloan Digital Sky Survey \citep{Yor00}
are much more restricted by 
the nonlinearity of clustering, they do provide a valuable data point
at an epoch where the dark energy is largest.

It is worth comparing the measurements from future redshift surveys to
those inferred from the observations of type Ia supernovae (hereafter
SNe) \citep{Riess98,Perlm99,Rie01,Ton03}.  
The SNe survey measure the luminosity
distance as a function of redshift, which in standard cosmologies is
equivalent to the angular diameter distance. While this requires an
additional derivative to extract $\w(z)$ relative to measures of $\hz$,
future SNe program such as the SNAP satellite could achieve extremely
good precision on distances at redshifts below 1.7.
While the cosmological implications of low-redshift
acoustic oscillations and SNe distances are partially degenerate,
the systematic errors will be completely different.

In summary, the baryon acoustic oscillations form a standard ruler that
can be measured through galaxy redshift surveys to yield $H(z)$ and $D_A(z)$
at a range of redshifts.  
The scale of the acoustic oscillations is expected to be
very robust to non-linear gravitational clustering, galaxy biasing, 
and redshift distortions, making this a potentially clean probe of
cosmography.  If we can show that the distance measurements can be made 
to sufficient precision, then acoustic oscillations will offer an
new and independent path to the quantification of dark energy.

\section{Methodology} \label{sec:Metho}
In this section, we present the methodology of constraining the dark
energy through distance measurements derived from surveys of galaxy clustering. 
To probe the time
evolution of the dark energy, we need galaxy power spectra at a variety of
redshifts.  We design surveys at six different redshift bins, ranging
from 0.3 to 3.  In this section, we present our methodology for
computing the statistical errors from these surveys and from our
ancillary data sets.  We do this using a Fisher matrix formalism
in a parameterized cosmological model.

\subsection{Statistical Error on the Power Spectrum } \label{subsec:sigmap}
To estimate errors on $\DA$ and $\hz$, we begin with the errors on the
power spectrum that result from a galaxy survey. 
Under Gaussian approximations,
the statistical errors are a combination of the limitations of 
the finite volume of the survey and the incomplete sampling of
the underlying density field.  These are known as sample variance 
and shot noise, respectively. At a single wavevector $\vec{k}$, 
the intrinsic statistical error associated with power
is the sum of the power and shot noise
\citep{FKP94,Teg97}
\begin{equation}
\frac{\sigma_{ P}}{{ P}}=
\frac{P+\frac{1}{n}}{P}
\label{eq:err0}
\end{equation} 
Here, $1/n$ is a white shot noise from the Poisson sampling of the density
field assuming that the comoving number density $n$ is constant in
position. If the shot noise term exceeds the true power, that is, when
$nP$ is less than unity, then shot noise will significantly 
compromise the measurement. Note that $nP$ depends on wavenumber.

However, when the survey volume is finite, the power at nearby wavevectors
is highly correlated, and one can think of discretizing the Fourier
modes of the density field into cells in Fourier space whose volume
is $(2\pi)^3/\Vsur$, where $\Vsur$ is the comoving survey volume.
Neglecting boundary effects, the statistical power of the survey
is well approximated by treating these cells as independent \citep{Teg97}.
If the survey volume is large enough that the discretization scale
is small compared to the regions of wavevector space over which the
power spectrum is constant, then we can estimate the bandpower 
as averaged over a finite volume in Fourier space.  We parameterize
this by the wavenumber range $\Delta k$ and the range $\Delta \mu$
of the cosine of
the angle between the wavevector and the line of sight.
The volume in Fourier space is simply $2\pi k^2 \Delta k\Delta\mu$
and the number of modes is $2\pi k^2 \Delta k\Delta\mu\Vsur/(2\pi)^3$. However, because the density field is real-valued, the Fourier modes $\vec{k}$ and $-\vec{k}$ are not independent, which reduces the number of independent modes by a factor of two.
The fractional error on the bandpower is then \citep{FKP94,Teg97}: 

\begin{equation}
\frac{\sigma_P}{P}=2 \pi \sqrt{\frac{2}{\Vsur k^2 \Delta k \Delta \mu}} \left( \frac{1+nP}{nP} \right)
\label{eq:err}
\end{equation}

where $P$ is the average comoving bandpower.
This fractional error on power
spectrum (eq.\ [\ref{eq:err}]) enters in Fisher matrix and will be
propagated to the errors on parameters which we want to calculate.

\subsection{The Fisher Information Matrix for Galaxy Redshift Surveys}
\label{subsec:Fisher}
Given the uncertainties of our observations, we now want to propagate
these errors to compute the precision of constraints on cosmological parameters.
The Fisher information matrix provides a useful method for doing
this \citep[see][for a review]{Teg97a}.
The method takes as input a set of observables and a parameterized theoretical
model to predict those observables.
We denote the parameters as $p_1,\ldots,p_N$.
The Fisher information matrix incorporates the likelihood function of the
observables to yield the minimum possible errors on an unbiased estimator of
a given parameter, given that the true value of the parameters are that of
a so-called fiducial model.  Mathematically, these minimum errors are simply the
square roots of the diagonal elements of the inverse of the Fisher matrix.

Assuming the likelihood function for the bandpowers of a galaxy
redshift survey to be Gaussian,
the Fisher matrix can be approximated as \citep{Teg97}
\begin{eqnarray}  
F_{ij}&=&\int_{\vec{k}_{\rm min}} ^ {\vec{k}_{\rm max}} \frac{\partial \ln P(\vec{k})}{\partial p_i} \frac{\partial \ln P(\vec{k})}{\partial p_j} \Veff(\vec{k}) \frac{d\vec{k}}{2(2 \pi)^3}  
\label{eq:Fij} \\
&=&\int_{-1}^{1} \int_{k_{\rm min}}^{\kmax}\frac{\partial \ln P(k,\mu)}{\partial p_i} \frac{\partial \ln P(k,\mu)}{\partial p_j} \Veff(k,\mu) \frac{2\pi k^2 dk d\mu}{2(2 \pi)^3}  \nonumber 
\end{eqnarray}
where the derivatives are evaluated at the parameter values of the fiducial model 
and $\Veff$ is the effective volume of the survey, given as 
\begin{eqnarray}
\Veff(k,\mu) &=& 
\int \left [ \frac{{n}(\vec{r})P(k,\mu)}{{n}(\vec{r})P(k,\mu)+1} \right ]^2 d\vec{r}  \nonumber \\
&=&\left [ \frac{{n}P(k,\mu)}{{n}P(k,\mu)+1} \right ]^2 \Vsur,
\label{eq:Veff} 
\end{eqnarray}
where the last equality holds only if the comoving number density $n$
is constant in position.
Here, 
$\mu = \vec{k}\cdot \hat{r}/k$, where
$\hat{r}$ is the unit vector along the line of sight and
$\vec{k}$ is the wave vector with norm $k=|\vec{k}|$.
Due to azimuthal symmetry around the line of sight, 
the power spectrum $P(\vec{k})$ depends only on $k$ and $\mu$,
but of course it has an implicit dependence on the cosmological
parameters $p_i$.
Equations (\ref{eq:Fij}) and (\ref{eq:Veff}) are not fully general,
as we have assumed a flat-sky approximation in which the survey box
is imagined to be far from the observer.  Given that the clustering
scales of interest will subtend small angles on the sky in all of
our designed surveys, this is an appropriate approximation.

We have not included information from all wavenumbers in our equation
(\ref{eq:Fij}).  Wavenumbers smaller than $k_{\rm min}$ or larger than $\kmax$
have been dropped.  We use $\kmax$ to exclude information from the 
non-linear regime, where our linear power spectra are inaccurate.  We 
adopt conservative values for $\kmax$ by requiring
$\sigma(R)=0.5$ at a corresponding $R = \pi/2k$.
At $z=0$, this sets $\kmax=0.1\ihMpc$, which is
consistent with the numerical simulations of
\citet{Meiksin99}
and noticeably smaller than that used by most published analysis of past
redshift surveys. The $\kmax$ values used for different redshift bins
are listed in Table \ref{tab:con}.
The maximum scale of the survey $k_{\rm min}$ has almost no effect
on the results; we adopt $k_{\rm min}= 0$.

In principle, the mapping from the observed galaxy separations to 
the physical separations and wavevectors depends upon the cosmological
functions $D_A(z)$ and $H(z)$, which are varying continuously across
the redshift range of the survey.  When doing an analysis of real data,
one would of course include this variation.  For our forecasts, however,
we opt to break the survey into a series of slabs in redshift, inside of 
which we treat the survey region as a fixed Euclidean geometry, with a 
constant $D_A$ and $H$ and a rectilinear division between the transverse
and radial directions.  This approximation is harmless as regards the
statistical power of the survey or the parameter degeneracies involved.
We use redshift bins that are narrow enough to finely sample the dark
energy behavior.

\subsection{Parameters} 
A Fisher matrix formalism relies upon a detailed parameterization of
its space of models.  The performance forecasts are only as realistic
as the generality of the permitted models.  For our forecasts, 
we proceed in two stages.  First, we define a very general parameterization
based on CDM cosmologies and assigning independent parameters to each
redshift bin.  This permits us to forecast cosmographical constraints
independent of any dark energy model.  Second, we introduce a smaller
set of parameters to describe dark energy by relating the distances
in different redshift bins.  This will allow us to combine many distance
measurements into constraints on a low-dimensional dark energy model.

\subsubsection{Cold Dark Matter Cosmography} \label{subsubsec:param}
We use a very general space of cold dark matter models. Our parameter
include the matter density ($\Oh$), baryon density ($\Ob$), matter
fraction($\Om$), the optical depth to reionization ($\tau$), the spectral
tilt ($n_s$), the tensor-to-scalar ratio ($T/S$), and the normalization
($\ln{A_S}^2$). Our fiducial model is $\Om = 0.35$, $h=0.65$, $\OL=0.65$,
$\Ok=0$, $\Ob = 0.021$, $\tau = 0.05$, tilt, $n_s = 1$, and $T/S = 0$.

We supplement this model with many additional parameters to describe
the behavior at each redshift.  
For the CMB, we include an unknown angular distance $\DAcmb$ to the 
last scattering surface at $z=1000$.
For each redshift survey bin, we add a parameter for the 
angular diameter distance ($\ln D_A$), the Hubble parameter
($\ln H$), the linear growth function ($\ln G$), the linear redshift
distortion ($\ln\beta$), and an unknown shot noise $P_{\rm shot}$.
With 5 additional parameters in each of 6 redshift bins, the total
number of parameters for the CMB and galaxy surveys is 38.
The fiducial values of these parameters are evaluated at the central
redshift of each slice, and the fiducial values of $\beta$ are computed from
the values of the bias as found from the fiducial values of the
observed galaxy clustering.

By keeping $\DA$, $H(z)$, and $G(z)$ as separate parameters at each redshift,
we have avoided any assumption thus far of a specific dark energy model.
The only cross-talk between the various distances and amplitudes occurs
through the parameters of $\Oh$, $\Ob$, and $n_s$ that set the shape of the galaxy
power spectrum.  In other words, a good constraint at one redshift  
implies nothing for another redshift because we have specified nothing
about the behavior of the distances as a function of redshift.

The unknown white shot noise $P_{\rm shot}$ is a shot noise in the
observed power spectrum at each redshift bin that remains even after
the conventional shot noise of inverse number density is subtracted
from the observed power spectrum.  These terms can arise from galaxy
clustering bias \citep{Sel00} even on large scales because they 
zero-lag terms in the correlation function, which are permitted in
the theories of local bias \citep{Coles93}.

The partial degeneracy between redshift distortions and cosmological
distortions requires care because the broadband aspects of the observed
power spectra are extremely well-constrained in these surveys.  If one 
knew the precise amplitude of the matter power spectrum at a given redshift, 
then one would know the bias to high precision.  This would yield 
the value of $\beta$, and knowing this, we could extract the cosmological
distortions from the quadrupole distortions of the observed power.
Unfortunately, we do not regard this as a robust cosmological test.
Non-linear redshift distortions are not well understood, particularly
in the context of poorly constrained bias models.  We seek to isolate
our measurement of the cosmological distortions from overly optimistic
assumptions about redshift distortions.  The unknown growth functions 
and shot noises aid in this separation; the latter contributes because
a white noise limits the localization of a power-law break in a smooth 
power spectrum.  We do not use the
recovered growth functions in our dark energy fits. We will return to this topic in \S~\ref{subsec:Ob0.005}

\subsubsection{From Cosmography to Dark Energy} \label{subsubsec:eos}
We next wish to define a more restricted parameterization for the 
study of dark energy.  We do this through a simple parameterization
for the equation of state $w(z)$.
The equation of state of a cosmological constant has $w=-1$ at all times,
whereas quintessence models have $w>-1$, generically with time dependence.
While the most important distinction of dark energy models would
be to decide whether $w=-1$ or not, we also want to develop methods
for tracking the time dependence.
As a simplest approach, we assumed a linear
equation of state in redshift (eq.\ [\ref{eq:wz}]).
\begin{equation}
\w(z)=\w_0+\w_1 z
\label{eq:wz}
\end{equation}
Our choice of parameters for a dark energy is $\Ox$ (eq.\
[\ref{eq:hzda}]), $\w_0$, and $\w_1$. 
Other choices for parameterizing the free function $w(z)$
have been explored in \citet{Teg01b}, \citet{Lin03a}, and \citet{Hut03}.

We used a variety of dark energy fiducial models in this paper.
The parameters of these models are listed in Table \ref{tab:model}.
We will focus most of our attention on a $\LCDM$ model with 
$\Ox=0.65$, $w_0=-1$, and $w_1=0$ and on a comparison model
(Model 2) with $w_0=-2/3$.  The primary difference between these
is that dark energy remains more important at higher redshift in 
the $w=-2/3$ model.
We consider four models with redshift-dependent equations of state.
All have $w_1>0$, so that dark energy emerges at higher redshift
than we would infer from $w$ today.  In detail, we truncate the
increase in $w$ at early times by setting $dw/dz=0$ at $z>2$ so
that the value of $w$ at $z>2$ is simply $w(2)$.  This is of minor
importance because the dark energy is subdominant at these high 
redshifts, but it is necessary to avoid dark energy domination
at early times.  Models 5 and 6 have $w<-1$ today, which is a
challenge to theory \citep[but see][]{Cal98}; we include these
simply to study the phenomenological differences.

Equation (\ref{eq:wz}) defines the equation of state today as the
parameter $w_0$.  Because the observations are all at higher redshift,
the errors on $w_0$ are misleadingly poor, because uncertainties in
$w_1$ allow the value today to vary around a well-measured value at
higher redshift.  Errors on $w$ at higher redshifts decrease to a 
minimum at a redshift $z_{\rm pivot}$, similar to the central redshift
of the observations, and then increase again.
For any choice of $z_{\rm pivot}$, we can recast the parameterization
in equation (\ref{eq:wz}) as 
\begin{equation}
w(z) = w_0 + w_1 z = \wpivot + w_1(z-z_{\rm pivot})
\end{equation}
At this redshift of minimum error, the covariance between $\wpivot$
and $w_1$ vanishes, so that the two parameters are statistically
independent.  $z_{\rm pivot}$ can be computed from the covariance
matrix of $w_0$ and $w_1$ via the method in Appendix A of \citet{Eht99}.

\subsection{Completion and Transformation of Fisher Matrices}
We must complete our formula (eq.~[\ref{eq:Fij}]) for 
the Fisher information matrix for galaxy surveys
by identifying the power spectrum for the corresponding redshift bin. 
$P(\vec{k})$ in equation (\ref{eq:Fij}) is a
three-dimensional galaxy redshift power spectrum, to be reduced
to two dimensions by symmetry.
When we reconstruct our measurements of galaxy redshifts and positions
using a particular reference cosmology, which differs from the true cosmology, 
the observed power spectrum is
\begin{equation}
P_{\rm obs}(k_{{\rm ref}\perp},k_{{\rm ref}\parallel}) 
=\frac {\DA_{\rm ref}^2 \times \hz}{\DA ^2\times \hz _{\rm ref}}P_{\rm true}(\kpe,\kpa) + P_{\rm shot}.
\label{eq:P}
\end{equation}
Here, $D_A$ and $H$ values in the reference cosmology are distinguished
by the subscript `ref', while those in the true cosmology have no subscript.
$k_\perp$ and $k_\parallel$ are the wavenumbers across and along
the line of sight in the true cosmology.  
These are related to the wavenumbers calculated assuming the reference cosmology, by 
$k_{{\rm ref}\perp} = k_\perp D_A(z)/D_A(z)_{\rm ref}$ and 
$k_{{\rm ref}\parallel} = k_\parallel H(z)_{\rm ref}/H(z)$.
The prefactor of distance ratios accounts for the difference in volume
between the two cosmologies.
We adopt the reference cosmology to be equal to our fiducial cosmology
for simplicity.

Next, the true cosmology must be constructed, included the redshift distortions.
We do this by scaling to $z=0$:
\begin{eqnarray}
&&P_{\rm obs}(k_{{\rm ref}\perp},k_{{\rm ref}\parallel}) 
=\frac {\DA _{\rm ref} ^2 \hz}{\DA ^2 \hz _{\rm ref}} b^2\;\left(1+\beta\frac{k^2_\parallel}{k^2_\perp + k^2_\parallel}\right)^2 \nonumber \\
&&\quad\quad\quad\times \left (\frac {G(z)}{G(z=0)}\right)^2 
P_{{\rm matter},z=0}(k)+P_{\rm shot}
\label{eq:P1}
\end{eqnarray} 
where the bias $b$ is
$\Om(z)^{0.6}/\beta(z)$. 
The normalization used to derive the power spectrum at z is,
\begin{equation}
P(\kn,z=1000)=A_S^2 \frac{\kn}{\kfid}\left(\frac{c}{H(1000)}\right)^4 
\label{eq:norm}
\end{equation}
where $\kfid = 0.025/\Mpc$ and $\kn^{-1} = 3000\Mpc$.
The actual power spectrum and derivatives with respect to various
parameters are reconstructed from equation (\ref{eq:P1}), using the numerical
methods and results at $z=0$ from \citet{Eht99}.
 
For the Fisher matrix of CMB, we assume errors for the Planck satellite
including polarization from \citet{Eht99}. With Planck, the fractional 
error on
$\Oh$ and $\Ob$ are 0.9\% and 0.6\%, respectively.  Together, these
more than suffice to calibrate the sound horizon to 1\%. 
The recovered error on the angular diameter distance to $z=1000$ is 0.2\%.

\begin{figure}[t]
\centerline{\epsfxsize=3.8in\epsffile{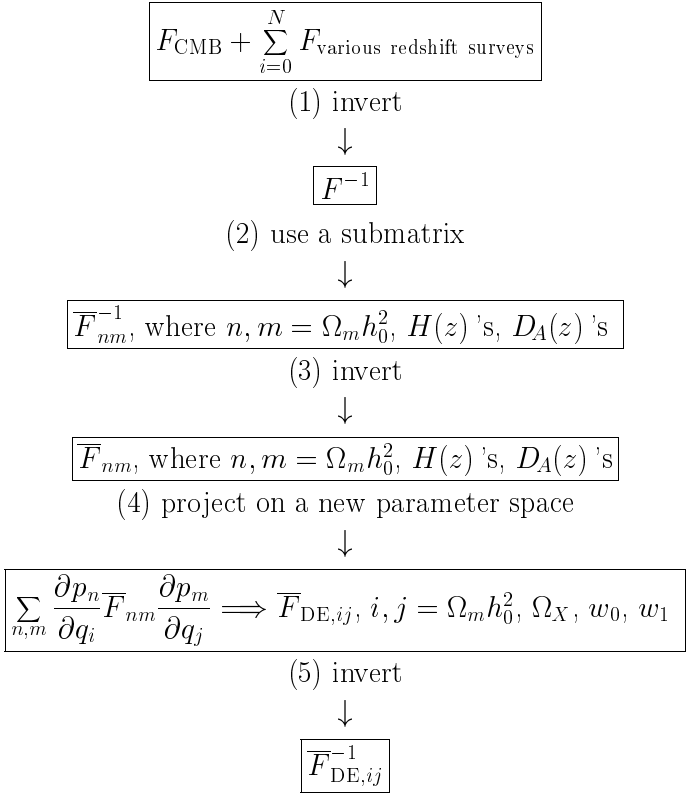}}
\caption{A flowchart of transformations of the Fisher matrices necessary
to produce forecasts for the distance and dark energy parameters.  }
\label{fig:tra}
\end{figure}

For the Fisher matrix of SNe, we introduce 16 redshift bins, at 0.05 and
at 0.3 to 1.7 by steps of 0.1, to represent the supernovae distance information.
We assign 1\% independent errors to each redshift point (i.e., 0.022 mag
error in distance modulus), with an overall 5\% uncertainty in the 
distance scale (since the SNe method by itself gives only a relative
distance measurement).  The appropriate covariance matrix is constructed
and then inverted to give the Fisher matrix.  In practice, the 
uncertainty in the distance scale is substantially reduced from the
5\% starting value by combination with the CMB, because the CMB's
measurement of $\Oh$ is combined with the SNe measurement of $\Om$
to yield the Hubble constant itself.

\begin{figure*}
\onecolumn
\plottwo{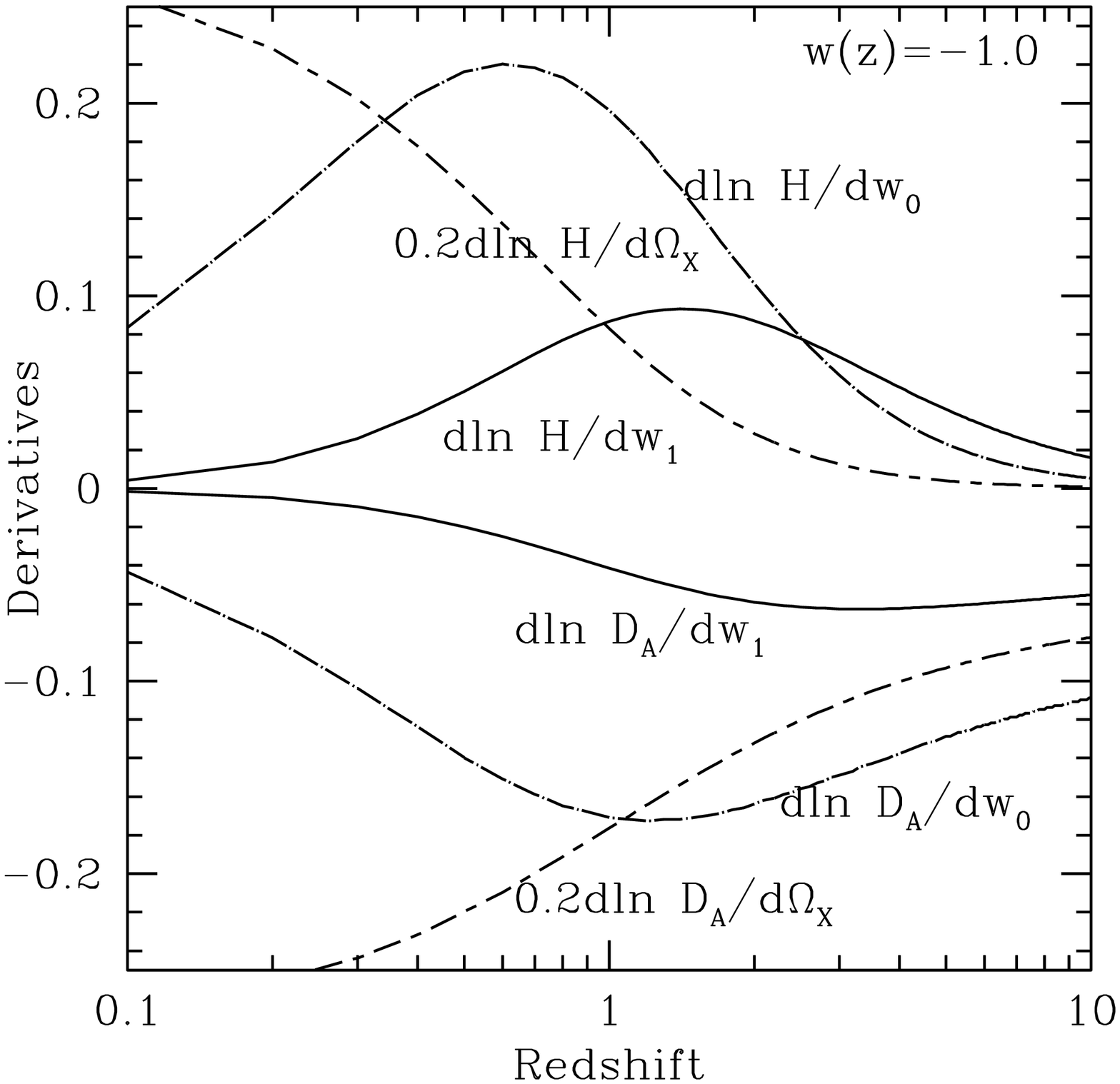}{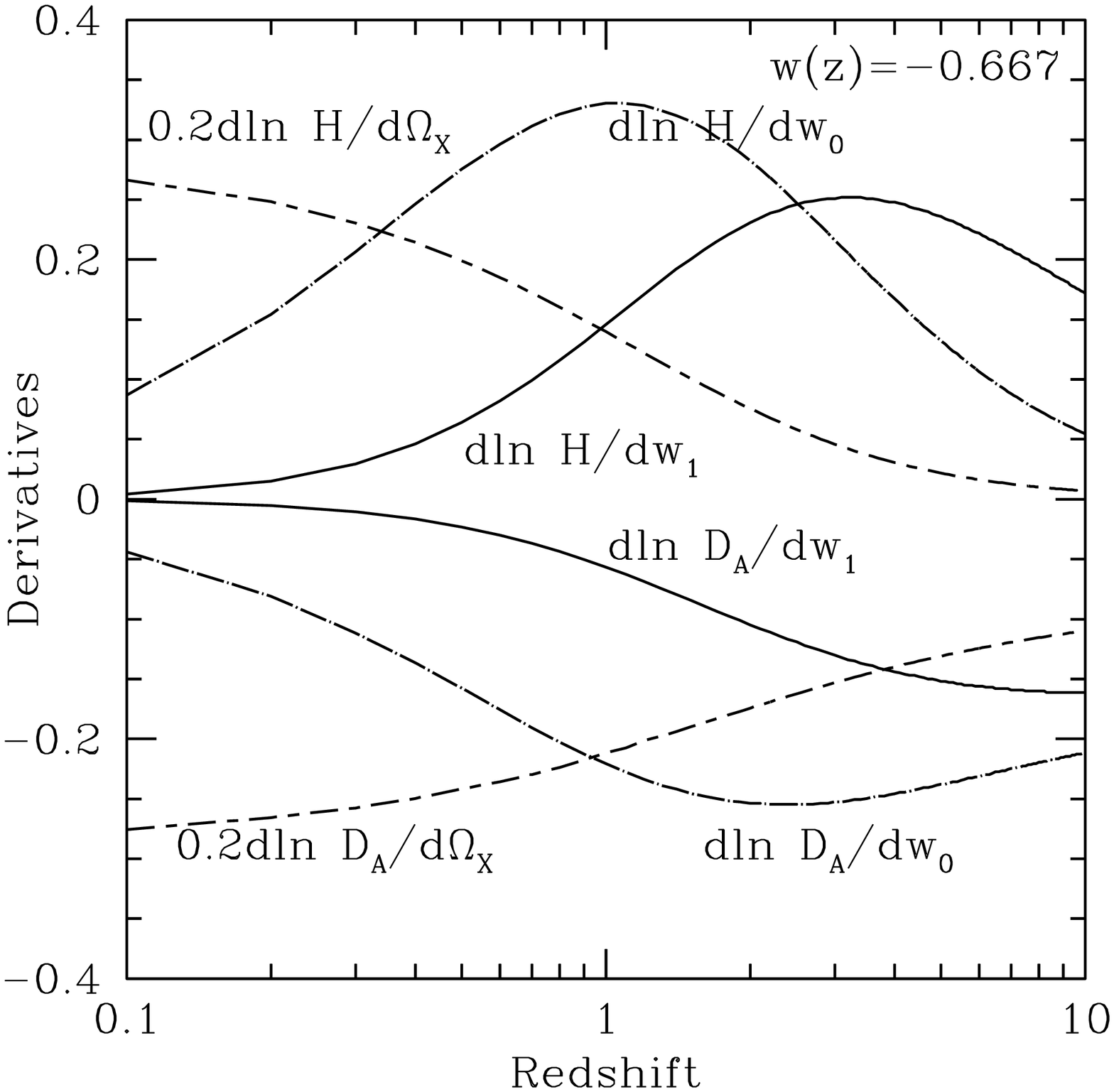}
\caption{ \label{fig:deriv}
Derivatives of angular diameter distance($D_A$) and Hubble
parameter($H$) with respect to $\Omega_X$, $w_0$, 
and $w_1$ with $\Oh$ being held fixed.
As these are partial derivatives, one should remember that 
two of the three parameters are held fixed as well in each case.
Notably, these are not the basis that leave the CMB anisotropies
unchanged.
The $\Omega_X$ parameter is equivalent to $\Omega_m$.
Left: $ w=-1.0$ ($\LCDM$). Right: $w=-0.667$ (Model 2). }
\twocolumn
\end{figure*}

Our SNe model was chosen to give similar performance to that of the 
proposed SNAP mission \citep{Ald02} but differs in fine detail from
that of the SNAP team.  One should note that our 16 redshift points
are statistically independent, so that with modest rebinning we are 
asserting better than 0.01 mag calibration between low and high redshift SNe.  
This is well beyond the current state of the art and is essentially the 
design goal of the SNAP mission.

Once the Fisher matrices for all the constituent data sets are set,
we must derive marginalized errors on $\DA$ and $\hz$ and eventually
on the dark energy parameters.  Figure \ref{fig:tra} shows the steps of
the procedure graphically.
To begin, the Fisher matrices are summed up and inverted.
The square roots of the diagonal terms of this inverse Fisher matrix are the
marginalized errors on parameters.
We marginalize over and remove all the parameters that are not concerned
with cosmography by taking a submatrix of the inverse Fisher matrix that
includes only the rows and columns for
$\Oh$, $\DAcmb$, and the $\hz$'s and $\DA$'s at all redshift bins.
This yields the covariance matrix for the cosmographical parameters.
Although this is an intermediate result, it is very useful because
it is independent of any dark matter model.

Next we project these errors through to the dark energy parameter space.
Because the dark energy model makes explicit predictions for the 
various distances, we are {\it not} marginalizing over parameters.
Rather, we are contracting the inverse of the covariance matrix, as
one would do in a multi-dimensional $\chi$-square analysis.
Hence, we invert the cosmographic covariance matrix to get a Fisher matrix
$\overline{F}$ and contract this with the set of derivatives between the 
the distances and the dark energy parameters 
($\Oh$, $\Ox$, $\w_0$, and $\w_1$).
\begin{eqnarray}
\overline{F}_{{\rm DE},ij}&=& \sum_{m,n} \frac{\partial p_n}{\partial q_i} \overline{F}_{nm}\frac{\partial p_m}{\partial q_j},
\label{eq:tran}
\end{eqnarray} 
where the $p_m$ are the distance parameters and the $q_i$ are the 
dark energy parameters.
By inverting this Fisher matrix, 
we attain marginalized errors of dark energy parameters.

Equation (\ref{eq:tran}) implies that the constraints on dark energy will
be a combination of how well $\DA$ and $\hz$ are estimated within a given
set of surveys and how effectively measurements of $\DA$ and $\hz$
can constrain dark energy.
Figure \ref{fig:deriv} shows the derivatives
of $D_A(z)$ and $H(z)$ with respect to the dark energy parameters.
The left panel is for $\LCDM$; the right panel is for Model 2 ($w=-2/3$). 
One should remember that these are partial derivatives, so that three of the parameters $\Oh$, $\Ox$, $w_0$, and $w_1$ are being
held fixed.
The derivatives with respect to $\w_0$ at fixed $\Oh$, $\Ox$
and $\w_1$ have larger amplitude than those to $\w_1$, meaning that
$D_A$ and $H$ place better constraints on $\w_0$ than on $\w_1$. 
Based on the positions of maximum
amplitudes, we expect that the information on $\w_1$ comes from higher
redshift than $\w_0$. It is interesting to note that while an advantage
of this acoustic oscillation method is to measure $\hz$, the peaks of
derivatives of $\hz$ are at lower redshift where, as we will find,
this method has poorer error bars. 
This tends to favor lower redshift probes such as SNe. 
It also implies that, improving error bars on $\w_1$ could
be done by changing the redshift survey conditions at higher redshifts,
that is, we may want to decrease error bars on $\hz$ over the range $z=1$ to 3
or on $\DA$ at $z \gtrsim 2$. 
Comparing $w=-2/3$ to $\LCDM$, one finds that the derivatives of
both $\DA$ and $\hz$ peak at higher redshift when $w$ is more positive.
This will favor the galaxy surveys at higher redshift.
Models 3 through 6 share this trend.

\subsection{Survey Design} \label{subsec:design}
We want to design redshift surveys that are optimized to derive $\DA$
and $\hz$ within accessible resources. Our requirement is that we
should be able to measure multiple acoustic peaks at various redshifts
with high precision.  In this section, we define two sets of baseline
surveys, with parameters in Table \ref{tab:con}; we will also consider
variations on these in \S~\ref{sec:RD}.

To constrain the scale of the acoustic peaks, we clearly need superb
precision in the power spectrum measurements.
Equation (\ref{eq:err}) shows that the errors $\sigma_P$ 
depend on the survey volume $\Vsur$ and on the number density $n$ of
objects in the survey.
Of course, $\Vsur$ and $n$ are limited by the available observational resources. 
If we assume that the observational resources scale with the total number 
of objects $N$, then at fixed $N$,
$\sigma_P/P$ has a minimum at $n=1/P$
\citep{Kaiser86}
at each wavenumber $k$. 
However, near this minimum, 
the performance $\sigma_P/P$ varies slowly, 
and a small deviation from the minimum
incurs little penalty. 
For example, using $nP=3$ or $nP=1/3$
increase the error by only about 15\%. 
With the relatively small dependence of error on $nP$ near the minimum, 
we suggest that $nP$ slightly larger than 1 is preferable for several reasons. 
First, larger $nP$ increase the signal-to-noise per pixel in the map. 
This enables computations beyond the power spectrum, e.g. for higher-order
correlations and non-Gaussianity.
Second, it avoids some complication from the non-Gaussianity of the shot 
noise itself.
Finally, it permits us to the survey into a few sub-samples based on 
galaxy properties or other criteria with less loss in signal-to-noise. 
This allows certain kinds of tests for systematic errors in the survey
and for additional science return from the study of type-dependent
galaxy bias.

On the other hand, it is possible that observation resources do not 
simply scale with
the number of objects. For example, the field of view, i.e. $\Vsur$,
may be more expensive than the number of spectroscopic targets. For a
fixed survey volume, the error bars improve monotonically as targets
are added, but the benefit saturates at $nP \gg 1$. 
For example, the error $\sigma_P/P$ with $nP=5$ is 1.7 times better
than that of $nP=1$ (at fixed volume), but only 20\% worse than that
of $nP=\infty$.
In reality, increased target density is not free: hence higher
number densities of objects require fainter objects (i.e., a deeper survey)
and hence longer exposure times. Fortunately, the range of the number
density we want is near the luminous tale of the luminosity functions,
where the source counts are quite steep, and so it is rather easy to
increase $n$ moderately above $1/P$.

We conclude that $nP\sim3$ is a good choice based on these considerations.  

An additional question is which wavenumber $k$ to use in calculating 
the value of $nP$.
We are primarily interested in higher acoustic peaks,
which occur around $k=0.2\ihMpc$.  The power at this wavenumber is about 2500$\sigma^2_{8,g}h^{-3}\Mpc^3$, where $\sigma^2_{8,g}$ is the
\textit{rms} overdensity of the galaxies in spheres of 8$\hMpc$ comoving
radius. This gives $n = 4\times10^{-4}\sigma^{-2}_{8,g}h^3\Mpc^{-3}$
for $nP = 1$.  This is considerably less than the
density of $L^*$ galaxies. Power is higher at smaller $k$, so smaller
densities would be optimal when measuring larger scales.

\begin{figure*}[t]
\epsscale{2}
\plottwo{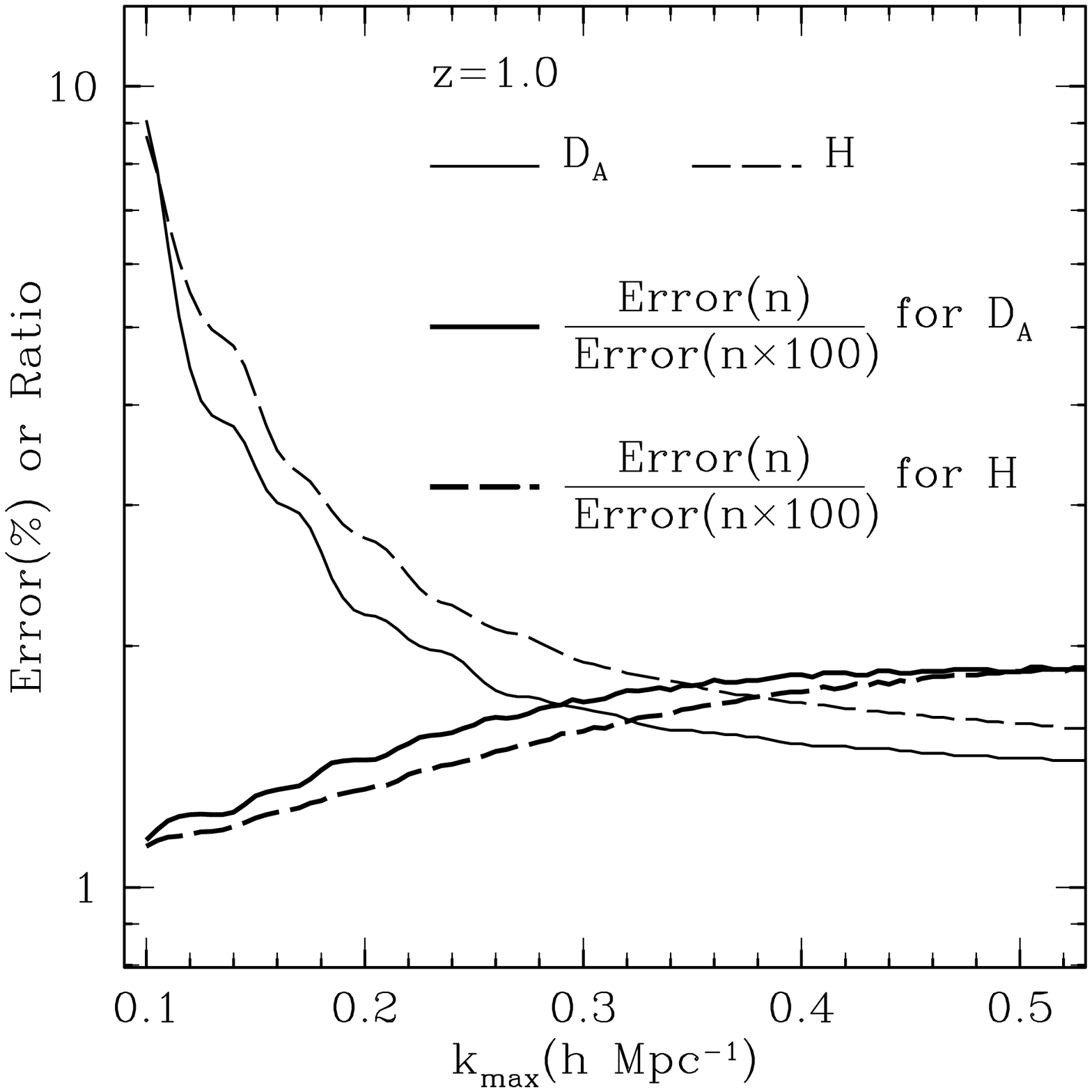}{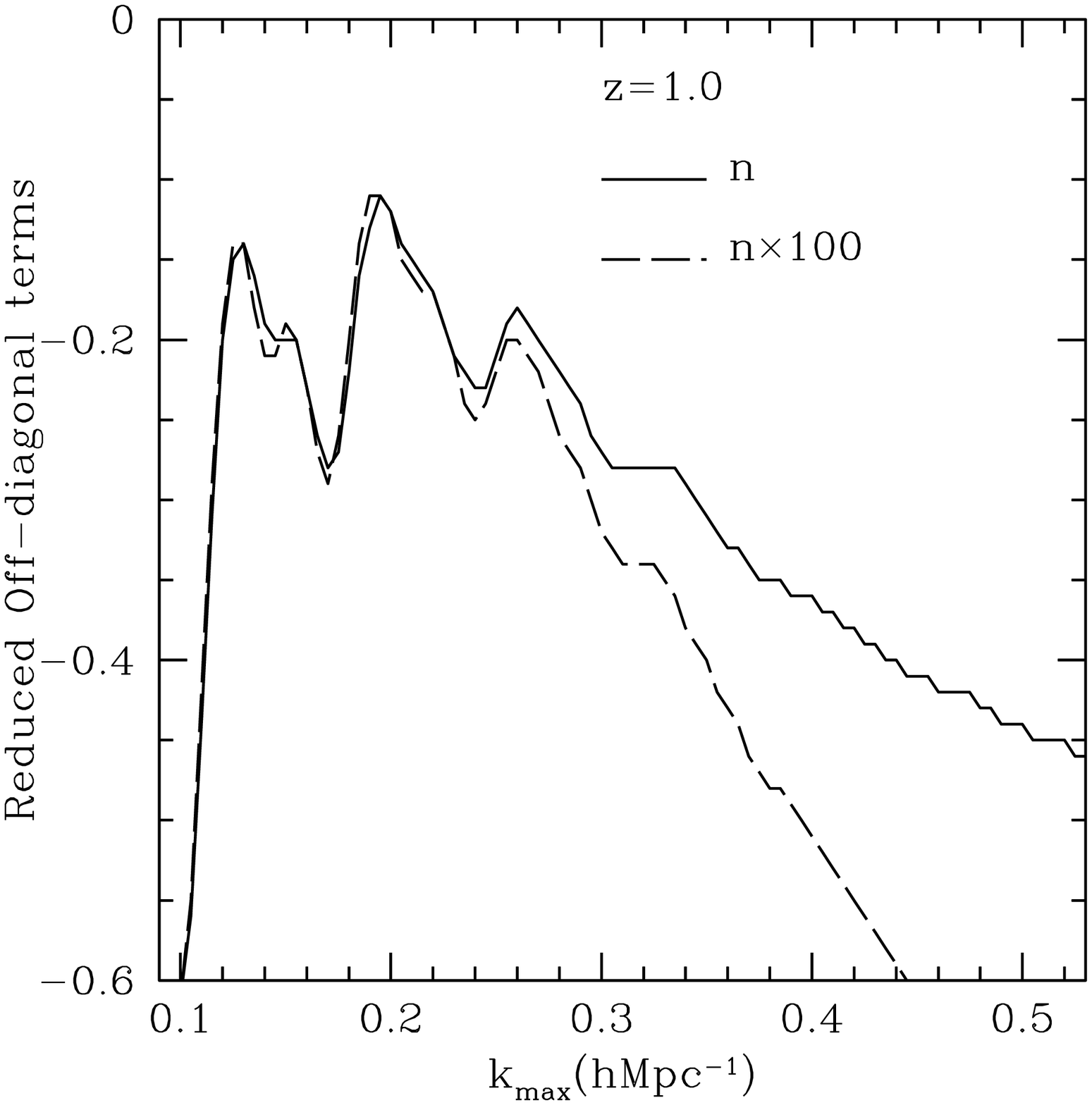}
\epsscale{1}
\caption{\label{fig:kmax}
Errors on $\DA$ and $\hz$ 
as a function of $k_{\rm max}$ and $n$ for $z=1$ data set. $n$
means the baseline number density in Table \ref{tab:con} 
(about $5\times10^{-4}h^3\Mpc^{-3}$), and $n\times100$
means 100 times the baseline number density. }
\end{figure*}

At $z \sim 3$, the obvious choice of galaxy targets are the 
Lyman-break galaxies \citep{Ste96}. 
$\sigma_{8,g}$ for these galaxies is measured to be about 1
\citep{Steidel98,Adel98}.
The corresponding bias is calculated using
\begin{equation} 
\sigma_{8,g} = b\sigma_{8,\rm mass} \sqrt{1+\frac{2\beta}{3}+\frac{\beta^2}{5}}
\label{eq:bias}
\end{equation}

assuming $\sigma_{8,\rm mass}$ of 0.9 for the matter distribution today 
and a linear redshift distortion effect \citep{Kaiser87}.
For the number density, $10^{-3}h^3\Mpc^{-3}$ is used so that $nP \approx 3$
at $k=0.2\ihMpc$. 
As an aside, a density $nP(0.2\ihMpc)>1$ 
is particularly valuable at $z=3$ because the non-linear
scale has receded to much smaller scales ($\kmax\approx0.5\ihMpc$!).  
To make full use of the 
survey at all linear scales, we need a larger $n$.
For our baseline survey, we adopt a 
total comoving volume 
of 0.5$h^{-3}{\rm Gpc^3}$, which gives enough resolution and
the precision to recover the first four acoustic peaks (Figure~\ref{fig:bumps}). 
At this redshift,
the comoving volume between $z=2.5$ and $z=3.5$ is 960$h^{-3}\Mpc^3$
per square arcminute, yielding a total survey field of 140 square
degrees. The areal number density is 1 galaxy per square arcminute,
similar to the depth of \citet{Steidel98}.     

At $z \sim 1$, the choice of galaxy target is less obvious. 
One could reasonably use
either giant ellipticals or luminous star-forming galaxies. Luminous
early-type galaxies have the advantage of high bias,
probably $\sigma_{8,g}>1$, and strong 4000\AA\ breaks,
but getting the redshift does require detecting this continuum break,
which takes longer integration time. Later-type galaxies may be less
biased, meaning we need a larger number density, but they have strong
3727\AA\ emission lines, which can often be identified because the line
is a doublet. For either case, we assume $\sigma_{8,g} = 1$ and $n =
5\times 10^{-4}h^3\Mpc^{-3}$. 
This makes $nP$ at $k \sim 0.2\ihMpc$ slightly bigger than 1, 
which means that $nP$ will be at our desired value for the meaty part
of the linear regime.
From $z=0.5$ to $z=1.3$, there is a comoving
volume of $480h^{-3}\Mpc^3$ per square arcminute, leading to a surface
density of 0.24 galaxies per square arcminute. We assumed total survey
field of 1000 square degree, chosen to sample a similar volume
to the Sloan Digital Sky Survey (SDSS) luminous red galaxy sample. 
The total number of galaxies
is 8.7$\times 10^5$.  To ensure sufficient resolution on the variations
of $\DA$ and $\hz$, we subdivide the $z\sim1$ survey into four
redshift bins centering at $z=$0.6, 0.8, 1.0, 1.2 with widths $\Delta
z=0.2$.  Hereafter, unless noted, the term `$z\sim1$ survey' 
designates the sum of these four redshift bins.

For the nearby universe, we adopt the parameters of the 
on-going SDSS luminous red galaxy survey 
\citep{Eis01}.
The survey volume for this sample is 1$h^{-3}{\rm Gpc^3}$, and the
comoving number density is $10^{-4}h^3\Mpc^{-3}$ at $z \approx0.3$. This
survey is included in all analyses in this paper because
it is well underway. 
We use $\sigma_{8,g}=1.8$ for these galaxies.

To resolve the oscillations along the line of sight at $k\approx 0.2\ihMpc$,
and thereby measure $H(z)$,
requires that the position of the galaxy along the line of sight be
well estimated.  As the crest-to-trough distance for this wavelength 
is only $15\hMpc$, we need redshifts with accuracy of $10^{-3}$ in
$1+z$.  We will return to this computation in \S~\ref{subsec:photoz},
but for now we note that this accuracy requires low-resolution
spectroscopy.  Photometric redshifts cannot recover $H(z)$ from
the acoustic oscillations.

\begin{figure*}
\epsscale{2}
\plottwo{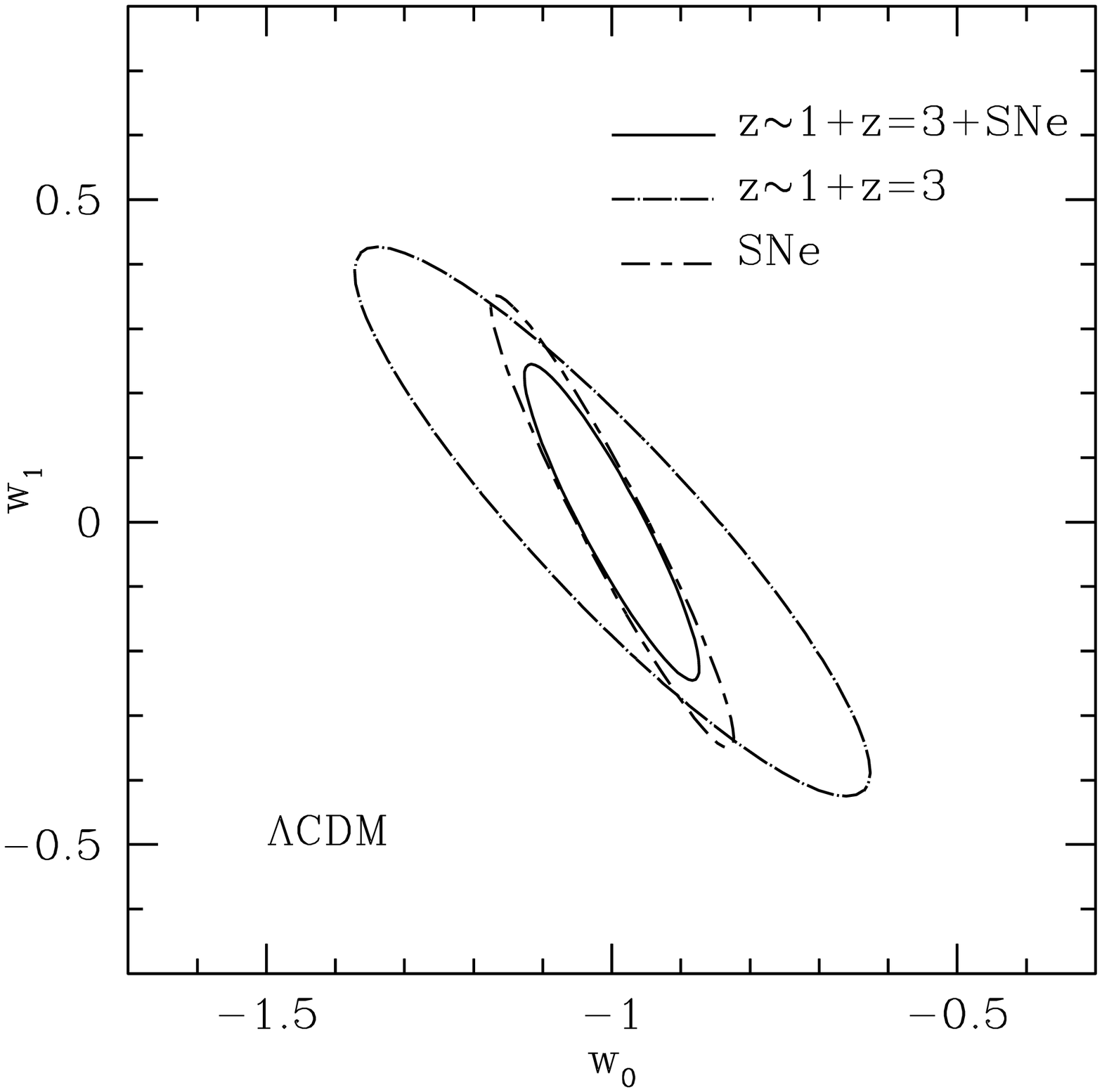}{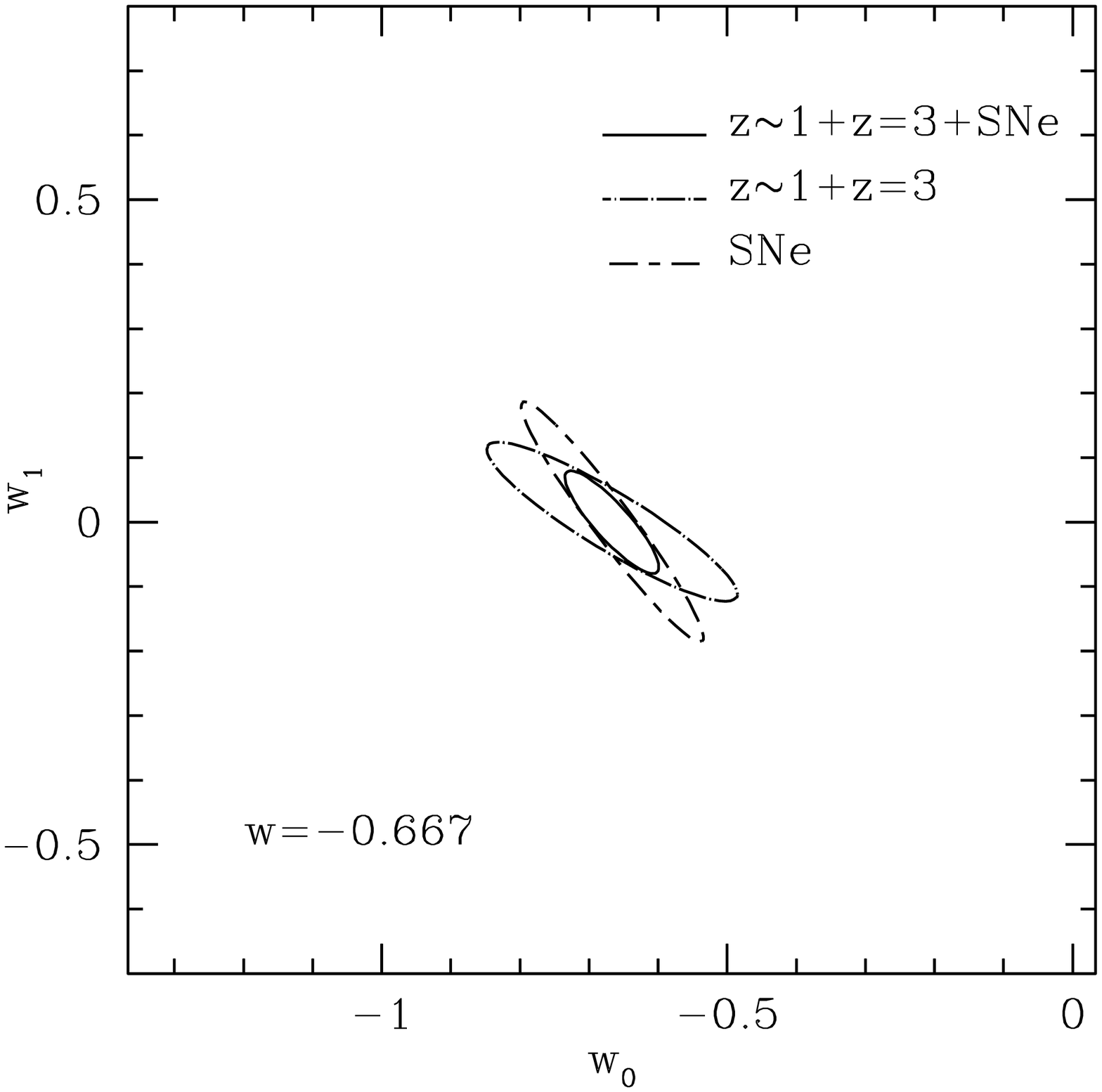}
\epsscale{1}
\caption{\label{fig:ellip}
Elliptical error regions on $w_0$ and $w_1$ for two different fiducial models. 
All other parameters have been marginalized over, and the contours
are for 68\% likelihood.
CMB and SDSS are included in all cases. }
\end{figure*}

\section{Results and Discussion}\label{sec:RD}
\subsection{Redshift surveys with SDSS and CMB} \label{subsec:result1}
We begin by presenting the results for cosmography from our baseline 
surveys.  
Table \ref{tab:dahz} lists the errors on $\DA$ and $\hz$ for
a combination of all the baseline redshift surveys and the CMB data. 
The errors
improve at higher redshift because of the smaller scale of the nonlinear
contamination. At $z=3$, the constraints are particularly good, better
than 2$\%$ on both quantities.
The errors on $\DA$
are generally smaller than those on $\hz$. This is simply because the
number of modes available in the two transverse directions is bigger
than the number of modes in the one line-of-sight direction. 

The reduced covariance matrix of the $\DA$ and $\hz$ values is shown in
Table \ref{tab:cor}. $\DA$ and $\hz$ at different redshifts are covariant
only through the uncertainty in the physical scale of the acoustic
oscillations. From the tiny non-diagonal terms between different redshift
bins in Table \ref{tab:cor}, we can see that the sound horizon scale is
very well determined. The non-diagonal elements of $\DA$ and $\hz$ in
the same redshift bin show that the degeneracy between the two is indeed
small as they are determined independently by the standard ruler test.

Most of the behavior in the errors can be explained as variations in
the non-linear cutoff scale $\kmax$ and in the survey sizes $\Vsur$. 
We explore this in Figure \ref{fig:kmax} by showing how the
performance at $z=1$ depends on $\kmax$. 
In the left panel of Figure \ref{fig:kmax}, we plot the errors
on $D_A$ and $H$ as functions of $\kmax$ for two values of the
number density $n$.
The drop from $\kmax=0.1\ihMpc$ to $\kmax=0.2\ihMpc$ dominates
the increase in performance from $z=0.3$ to $z=1.2$.

The errors on the distances flatten at around $\kmax \sim 0.25 \ihMpc$,
implying a saturation of the information from the locations of
baryonic acoustic peaks.  This is easily understood as the drop in 
contrast of the higher harmonics because of Silk damping.
Beyond this wave number, the errors slowly decrease with more efficiency
for $\hz$. This slight increase in information seems to be due to the
Alcock-Paczynski effect reappearing as the deviation the power
spectrum from a pure power-law is revealed by the increasing range
of wavenumbers in the survey.
 
The oscillatory behavior versus $\kmax$ shown in Figure \ref{fig:kmax} is
due to the oscillatory derivatives of the power spectrum with respect to
dilations in the distance scales. When $\kmax$ is close to the nodes
of power spectrum, the derivative $d\ln P/d\ln k$ has a local maximum
and the survey can better distinguish the differing cosmologies. 
The right panel of Figure \ref{fig:kmax} shows the covariance between
the uncertainties in $\DA$ and $\hz$.  These show a similar dependence
on 
$d\ln P/d\ln k$ but with a phase offset. 
When the performance improves suddenly, the ability to separate
the two distances has a local maximum.
Thus, the decrease of the non-diagonal
term at $z=1$ and $1.2$ in Table \ref{tab:cor} is simply because $\kmax$ has
shifted to be near one of the maxima of the plot in the right panel of
Figure \ref{fig:kmax}.  
The increasing covariance between $\DA$ and $\hz$ at very large $\kmax$
is another signature of the Alcock-Paczynski effect in the broadband
power that eventually intrudes.

Figure \ref{fig:kmax} also shows the degradation of performance caused
by shot noise.  We generate results with essentially zero shot noise by
increasing the galaxy number density by a factor of 100. 
This reveals the bare effect of $\kmax$ variations; with the baseline
surveys, the power spectrum errors at large $k$ are somewhat degraded
by increasing shot noise. 
The left panel of Figure \ref{fig:kmax} displays the ratio of 
performance in the two cases.  For $k\approx 0.2\ihMpc$, the 
degradation due to shot noise is less than a factor of 1.5, as
expected.  However, at large $k$, the effect is a full factor of two.
Improved performance at large $k$ increases the strength of the 
Alcock-Paczynski effect, as shown by the even larger covariance
in the high density case in the right panel.

\begin{figure*}
\epsscale{2}
\plottwo{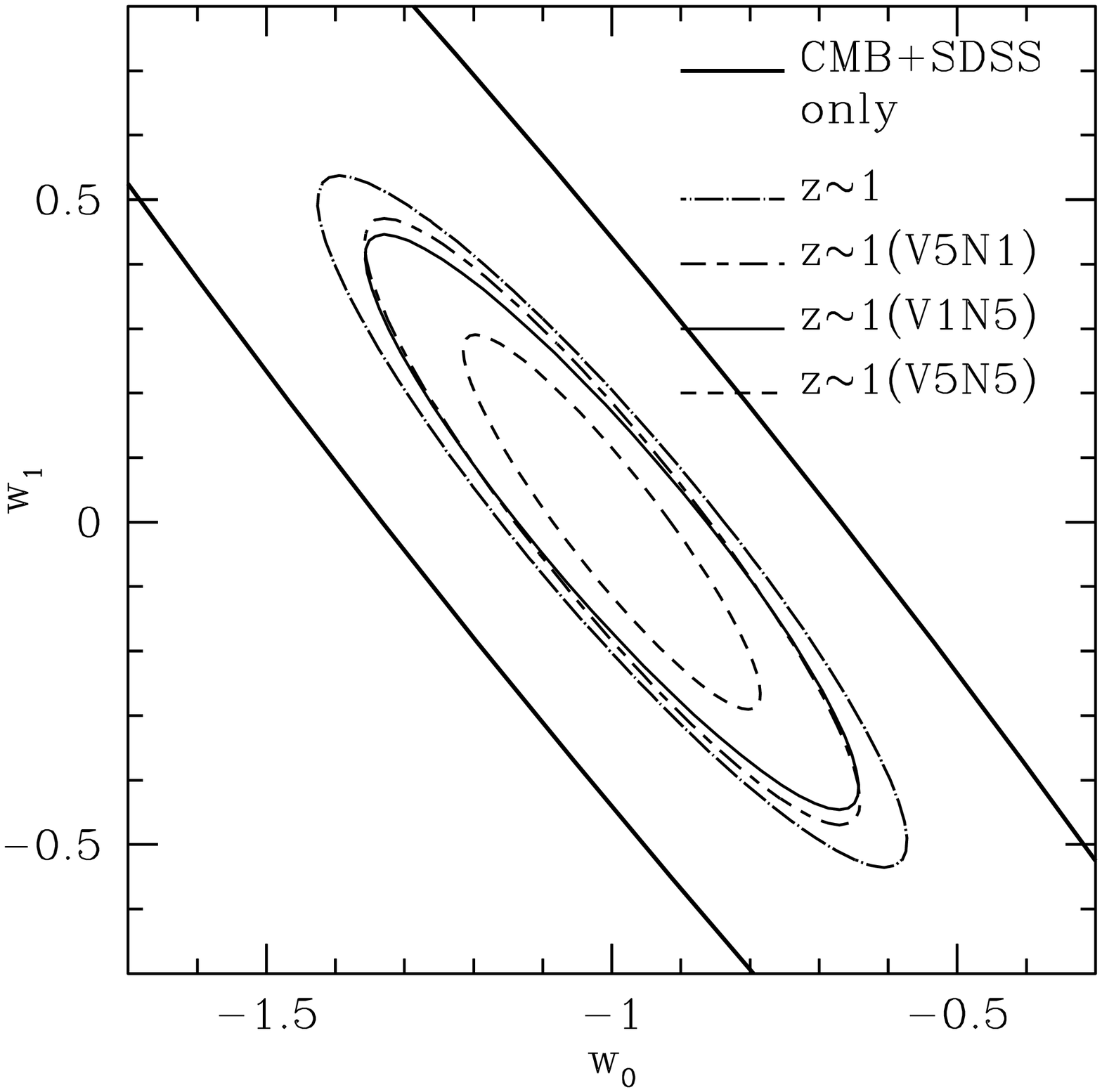}{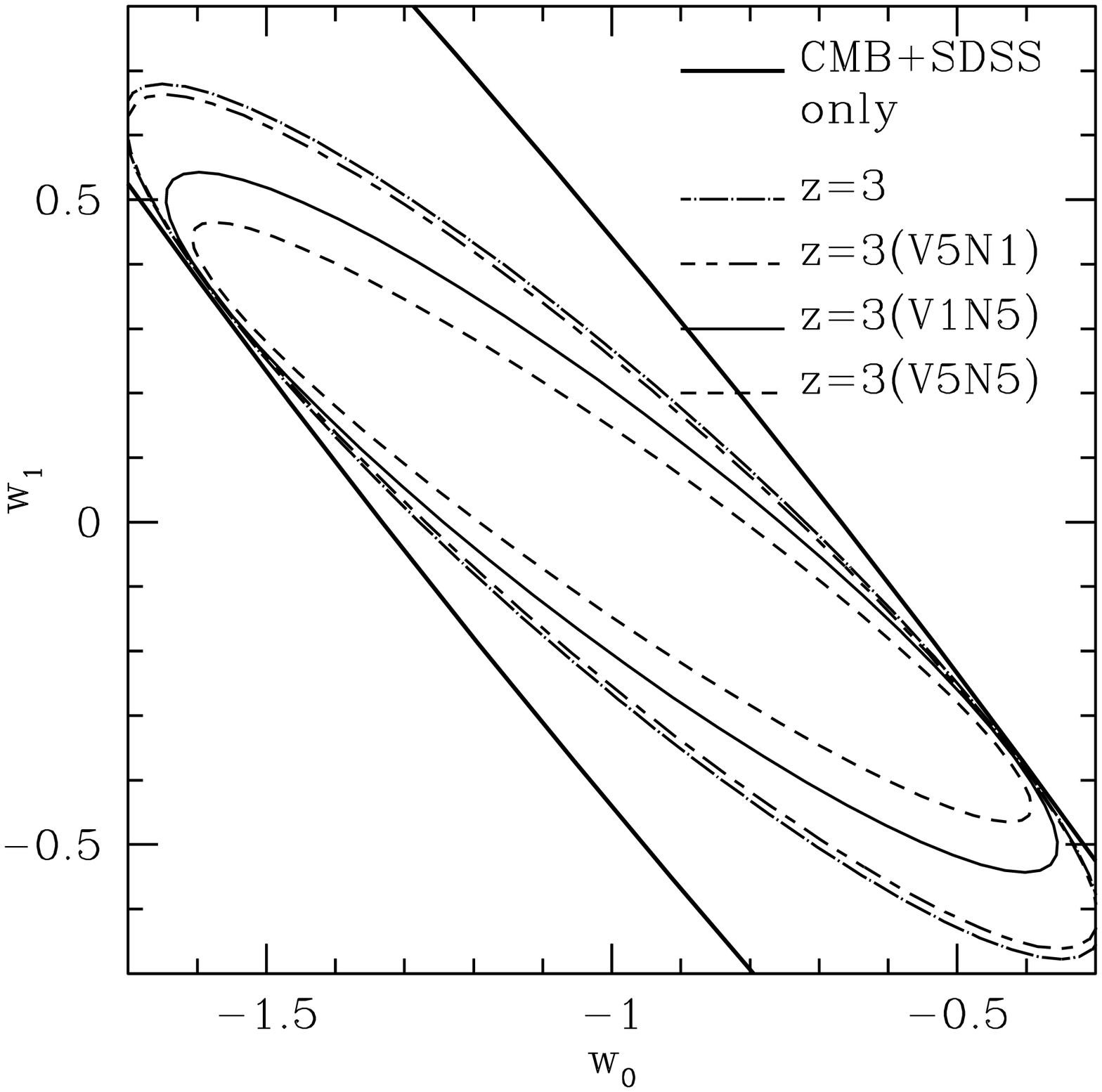}
\epsscale{1}
\caption{\label{fig:conellip}
Elliptical error regions on $w_0$ and $w_1$ as a function on survey parameters ($\Lambda CDM$). 
All other parameters have been marginalized over, and the contours
are for 68\% likelihood.
CMB and SDSS are included in all cases. V5N1 means 5 times the baseline
survey volume with 5 times smaller number density. V1N5 means 5 times
the baseline number density. V5N5 means 5 times the baseline survey
volume with the baseline number density. The ellipse with no notation
on survey parameters means the baseline survey parameters. For SDSS,
the baseline survey parameters was used in all cases.}
\end{figure*}

We next project the errors from the baseline surveys through to 
constraints on the dark energy parameters.
Table \ref{tab:resw} shows the performance on
dark energy parameters using fiducial model 1 ($\LCDM$).
With all redshift surveys combined with CMB and SDSS data,
we can achieve a precision of 0.037 on $\Ox$, 0.25 on $\w(z=0)$, 0.10
on $\w(z=0.8)$, and 0.28 on $\w_1$. For the $\LCDM$ model, as well as
for  Model 2 in Table \ref{tab:model}, we did not
clip $\w_1$ for $z > z_t$ with $z_t=2$. Clipping $\w_1$ for $z>z_t$
in $\LCDM$ increases the errors by a factor of 1.2.

In these calculations, we assumed not only that the errors on the
distances were Gaussian but that these generate a Gaussian likelihood
function for the dark energy parameters.  This is appropriate for
well-constrained parameters such as distances, $\Ox$, and $\Oh$, but
may be incorrect for $\w_0$ and $\w_1$.  We repeated our analysis
with a more complete likelihood calculation, in which the likelihood
at each point in $w_0$-$w_1$ space was computed assuming a 
(more appropriate) Gaussian likelihood in the other parameters.  
The result is the likelihood function in $w_0$ and $w_1$ with the 
other parameters marginalized out.
The resulting likelihood contours were not ellipsoids, of course,
and were slightly bent and offset.  However, the extent and slope
of the contours were excellent matches to the Gaussian ellipsoids.
We therefore conclude that the Gaussian analysis gives a reasonable
estimate of the dark energy performance and is sufficient
for comparing different combinations of surveys.

Constraints on dark energy fiducial models 2 through 6
are presented on Table \ref{tab:w}. Some of these non-$\LCDM$ models have
significantly improved performance on dark energy parameters.
In particular, models 2 ($w=-2/3$) and 3 ($w_0=-2/3$, $w_1=1/6$)
have superb performance, with constraints on $w_1$ reaching 9\%.
Models 4 and 6 are also better than $\LCDM$.
Not surprisingly, these improvements correlate directly with the value 
of $w$ at intermediate redshifts and hence with the amount of dark energy
that remains at higher redshift.
Most of improvements are keyed by the measurement of $\hz$ and $\DA$ 
at higher redshifts. This is reflected in the systematic increase 
of $z_{\rm pivot}$ in the cases of improved performance.

\subsection{Incorporation of Supernovae Data}
We next combine these redshift surveys with the SNe data set.
The lower four rows of Table \ref{tab:resw} show the error on dark energy
parameters with the SNe survey. To begin, SNe data with only CMB and SDSS
data yields impressive performance.
$\Ox$ and $w_0$ are well constrained, 
and the error on $\w_1$ is 0.23, slightly better than
what the redshift surveys produce. When we combine the SNe data with the
galaxy redshift surveys, the $\w_1$ error improves to 0.16. With the SNe 
and CMB data,
the inclusion or exclusion of SDSS does not change the result much 
because of the relatively large uncertainty both in $\hz$ and $\DA$ from
SDSS as compared to the performance of SNe; most of the information 
in the survey is superceded by SNe data.

Figure \ref{fig:ellip} shows the constraints in the $\w_0$-$\w_1$
plane as error ellipses, marginalizing over all other parameters. The
left panel shows the $\LCDM$ model and the right panel shows Model 2
($\w=-2/3$) as a comparison to $\LCDM$. We see the difference in the
directions of the two ellipses: SNe with CMB and SDSS, and redshift
surveys with CMB and SDSS. 
The set with SNe shows a tight constraint
especially in $\w_0$ direction, and the improvement of the constraint on
$\w_1$ by redshift surveys.
By comparing two models, we can easily see that Model 2 allows much
better constraints on parameters than $\LCDM$ and favors redshift surveys
more. The redshift survey data is now comparable to the capability of SNe:
in $\w_1$, redshift survey data achieve 0.08, SNe survey data produces
0.12, and together the data sets produce 0.05.

The supernova data has superb precision for $z<1.7$ in $\DA$
and gives excellent constraints on the shape of distance-redshift
relation. 
Our baseline redshift surveys, on the other hand, have larger error 
bars than SNe for
$z\leq1.2$, but they have an advantage of having a distance-redshift
data point at very high redshift ($z=3$) and measuring $\hz$ in all
redshift bins. In the $\LCDM$ fiducial model, the contributions to
$\w_1$ by $\hz$ measurements and $\DA$ from the  $z=3$ redshift survey
are slightly less useful than the good precision of $\DA$ from SNe at
lower redshifts (see Figure \ref{fig:deriv}). On the other hand, dark
energy models with more positive $\w$ have larger signatures at higher
redshift.  This is good for both data sets, but helps the redshift surveys
more.

\subsection{Variation with Redshift Survey Parameters}
We next show how performance varies with survey parameters such as the
total number of galaxies $N$ and survey volume ($\Vsur$). 
We present variations in $\Vsur$ and $N$ by factors of five in 
Table \ref{tab:dahz1}.  Because the cosmographic performance in each redshift
survey is essentially independent, one can interpret this table
as varying each survey independently.  The SDSS and CMB data are
unchanged in all cases.
From Table \ref{tab:dahz1}, we can see that performance at $z=3$
is more sensitive to the increase in $N$ at fixed $\Vsur$ (i.e.
higher target density) than for the reverse. For the $z\sim1$ surveys, the effect of increasing the number density
is slightly larger for $z=1.0$ and 1.2 bins and is more efficient for
$\DA$ than for $\hz$.  Increasing $\Vsur$ was more effective for $z=0.6$
and 0.8 bins with a general trend of being more efficient for $\hz$
than for $\DA$. This agrees with the result from Table \ref{tab:con} that
the $nP$'s of $z=1.0$ and 1.2 are somewhat less than those of
$z=0.6$ and $z=0.8$. The preference to $\hz$ when decreasing the number
density is due to an increased contribution from wave vectors
along the line of sight, which suffer less shot noise degradation due
to their enhanced amplitude from redshift distortions.

The projected errors on the dark energy parameters under these various
survey parameters are presented in Table \ref{tab:surveycon}. The results
are consistent with the changes in the errors on $\DA$ and $\hz$.
The graphical illustration of these errors are shown by error ellipses in
Figure \ref{fig:conellip}.  For this figure, the surveys at $z\sim1$
(left panel) and $z=3$ (right panel) are used separately so that one
can see the individual scalings.
As one would expect, larger surveys provide better constraints.
The slopes of the major axes are an indicator of the typical redshift
$z_{\rm pivot}$ of the data.  
The twisting of the major axis direction in the $z=3$ case is a 
visual sign that larger $z=3$ surveys pull $z_{\rm pivot}$ to
be higher than the CMB and SDSS data would yield by themselves.

As regards the survey size, increasing $\Vsur$ at fixed number
density causes the performance to scale nearly as the square root of
$\Vsur$.  In detail, the results fall slightly short of this scaling
because the SDSS and CMB data are not be similarly scaled.  For
factor of 5 increases in the $z=3$ survey, one begins to see the
limitations of the CMB calibration of the sound horizon.

When combined with the SNe data, it is more valuable to improve the
$z=3$ survey than the $z\sim 1$ survey. Increasing $\Vsur$ by a factor of five (V5N5)
for both surveys improve the errors on $w_1$ by a factor of 1.6, 
increasing the $z\sim1$ surveys alone yields a 1.3 improvement,
whereas increasing $z=3$ alone improves by a factor of 1.4. Pictorially, this is because
the $z=3$ constraint ellipsoids for dark energy fall at more of an 
angle as compared to the SNe ellipsoids than do the $z\sim1$ ellipsoids.
Physically, it is more advantageous to widen the redshift range of 
the measurements,
especially because the SNe data has somewhat higher precision than the
$z\sim1$ redshift survey constraints on $\DA$.

As mentioned in \S~\ref{subsec:design}, adjusting the survey volume $\Vsur$ 
while holding the total number of targets fixed has an optimum point
for the measurement of the power spectrum at $nP=1$.  We therefore expect
that this trend would extend to performance on dark energy parameters.
Indeed, we find that slightly larger surveys (e.g., a factor of 2--3) 
do give small improvements
and that much larger surveys give steadily worse performance. Again, this is
exactly as we expected with our choice to aim for $nP\approx 3$.
True optimization of course requires detailed knowledge of the survey
instrument, the source population, the possible systematic errors, 
and the other science goals of the survey.

\subsection{Baryonic Oscillations versus Broadband Constraints}
\label{subsec:Ob0.005}   
To this point, we have discussed the baryonic oscillations as a distinct signature from which to infer cosmological distances. Although these features are essential, the Fisher matrix we calculate includes additional contributions such as the overall broadband shape of the power spectrum. In this section, we briefly assess the amount of information on distances from baryonic oscillations apart from other contributions.

To single out the non-baryonic contribution, we repeat our calculations with a fiducial model with ten times lower baryon fraction ($\Omega_b=0.005$), thereby removing the acoustic oscillations from the power spectrum. Overall, the errors on $\DA$ and $\hz$ increase by a factor of 2 to 3, with more increase in the $z\sim1$ set and more increase in $\DA$ than $\hz$, making the magnitudes of $\sigma_\DA$ and $\sigma_\hz$ nearly equal. The reduced correlation coefficient between $\DA$ and $\hz$ at the same redshift is about $-0.8$, indicating a strong correlation. This covariance and the more equal precisions imply that the \citet{AP79} test (hereafter AP test) is playing a significant role in constraining distances in the low baryon case. The combination $D_A H$ is well constrained, whereas the separate values of $\DA$ and $\hz$ are constrained only by the broadband shape of the power spectrum. 

The AP effect can isolate cosmological distortions in two ways. First, when the power spectrum has a preferred scale---and any deviation from a power law will suffice---we can measure the cosmological distortion $D_A H$ by requiring that scale to be isotropic. The values of $\DA$ and $\hz$ can be determined separately only if the preferred scale is known, for example, from CMB data. Second, one can attempt to separate the cosmological distortions from the redshift distortions by the angular dependence of the power spectrum at a given $k$. When the redshift distortion is weak ($\beta \approx 0$), the two distortions have identical angular signatures, both quadratic in $\mu$, and hence are indistinguishable. However, for larger $\beta$, both distortions take on more complicated forms that lift the degeneracy in principle.

Because in our analysis the shape of the power spectrum is known from the CMB data, the first mode of the AP effect cannot produce the strong covariance between $\DA$ and $\hz$ that we find in the $\Omega_b=0.005$ case. Hence the degeneracy between the redshift distortions and the cosmological distortions must be angularly broken (i.e. the second mode of the AP effect). To test this, we introduce a strong degeneracy between $D_A H$ and $\beta$ by using $\beta \approx 0$. For numerical reasons, we decrease the fiducial $\beta$'s 30-fold. We apply these lower $\beta$'s only to the computations of the derivatives; the original $\beta$'s are retained in computing $\Veff$ so that the weighting of the radial and tangential modes is unchanged. The upper two rows in Table \ref{tab:b0.005} show the results with $\Omega_b=0.005$. With negligible $\beta$'s, the errors increase by $15\%\sim35\%$ compared to the case with the normal $\beta$'s, with more increase for $\DA$ than $\hz$ and more increase in the $z\sim1$ set, which has larger $\beta$ than the others. The reduced correlation coefficients decrease to $\sim -0.3$, supporting the interpretation that the AP effect has been removed and the remaining constraints are due to the shape of the broadband power spectrum. 

We next apply the same method for $\Omega_b=0.05$ case so as to enforce degeneracy between the cosmological and redshift distortions. The lower half of Table \ref{tab:b0.005} shows the errors on distances in this case. The comparison between $\Omega_b=0.05$ case and $\Omega_b=0.005$ case in Table \ref{tab:b0.005} shows that the broadband spectrum is a minor effect compared to the baryonic oscillations. Comparing these results to the previous results in Table \ref{tab:dahz} shows that the performance from the baryonic oscillations will decrease by $10\sim50 \%$ if we assume that we do not know the behavior of the redshift distortions very well. 

To summarize, in the absence of baryonic oscillations, the AP effect is capable of constraining the combined quantity $D_A H$ very well provided that the shape of the redshift distortions is relatively well-known \citep{Bal95,Heavens97,Hatton99,Taylor01,Mat02,Mats03}. However, it is the baryonic oscillations that separate $\DA$ and $\hz$ most effectively and provide precision constraints regardless of the amount of information on the redshift distortions.

\begin{figure}[t]
\plotone{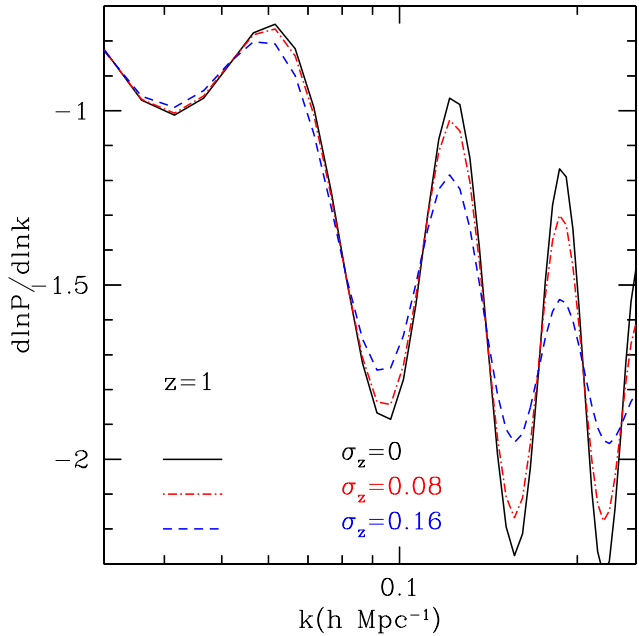}
\caption{
The derivative $d\ln P/d\ln k$ as a function of wavenumber for
three different values of the redshift uncertainty.
Larger uncertainties cause line-of-sight projections that smear
the narrow acoustic oscillations and impede the detectability of
cosmological distortions. 
$\sigma_z=0.0$ represents spectroscopic redshift error. $\sigma_z=0.08$ and
$\sigma_z=0.16$ corresponds to $4\%$ and $8\%$  of $\sigma_z/(1+z)$
at $z=1$.}
\label{fig:P2K}
\end{figure}

\subsection{Photometric Redshift Surveys}
\label{subsec:photoz}
With the advent of deep wide-field multi-color imaging surveys, it is
natural to ask whether photometric redshifts can be used for
studies of the acoustic oscillations.  In this section, we will
study how uncertainties in the galaxy redshifts affect our results.
There are two basic lessons.  First, recovering the Hubble parameter
$\hz$ requires measuring clustering on fairly small scales along
the line of sight, such redshift precision substantially better
than 1\% is needed.  Second, redshift slices selected with photometric
redshifts can be sufficiently thin that the acoustic oscillations
survive in the angular power spectrum.  This means that one can
measure $\DA$ with these surveys, albeit with worse precision per
unit survey sky coverage.
Hence, photometric redshift surveys lose the advantage of the 
acoustic oscillations to measure $\hz$ directly but could measure
$\DA$ if one has a large enough survey.
The idea of using transverse clustering to probe dark energy
was analyzed in the weak lensing context by \citet{Cooray01}.

When redshifts are uncertain, one is smearing together clustering
at multiple distances along the line of sight.  Our first task
is to consider whether the acoustic oscillations, being narrow
features in Fourier space, can survive this projection.  The controlling
effect is the variation in the comoving angular diameter distance
across the range of redshift uncertainty.  This can be addressed
with Limber's equation \citep{Lim53,Bau94}.  We model the redshift
distribution as a Gaussian of width $\sigma_z$ and consider the
effects on the angular power spectrum.  This is shown in Figure
\ref{fig:P2K}, where we plot the derivative $d\ln P/d\ln k$ that
controls the measurement of cosmological distortions for 3 different
values of $\sigma_z$.  We adopt a $z=1$ slice and consider 0\%, 4\%,
and 8\% uncertainties (1--$\sigma$) in $1+z$.  One can see that the oscillation
pattern is essentially intact at 4\% but substantially degraded at
8\%.  In detail, we estimate that the errors on $\DA$ would be 
increased by 13\% for the 4\% case and 54\% for the 8\% case.
The effects at $z=3$ are slightly more forgiving, despite the 
higher $\kmax$ and hence narrower features, because the derivative
of $\DA$ versus $z$ is slightly less.  
We therefore conclude that photometric redshift errors of 
4\% in $1+z$ (1--$\sigma$) are sufficient to preserve the acoustic
oscillations for the measurement of $\DA$ at $z\gtrsim0.5$.

Having found that the transverse power spectrum is not affected by
reasonable projections, we next include the redshift uncertainty in our Fisher
matrix formalism.  We do this by retaining the Euclidean approximation,
i.e. treating the survey as a box of fixed $D_A$ and $H$, but smearing
the radial position by a Gaussian uncertainty.  If the line-of-sight
comoving position $r_z$ is convolved with an uncertainty of the form 
$\exp[-(\Delta r_z)^2/2\sigma_r^2]$ with an uncertainty $\sigma_r$,
then the Fourier transform of the density field will simply be diminished
by the transform of this kernal: 
$\delta_{\vec{k}} \propto \exp(-\kpa^2 \sigma_r^2/2)$. The observed 
power spectrum is then (L.~Hui, private communication)
\begin{equation}\label{eq:onlyda}
P(\vec{k})=P_{\rm obs}(\vec{k}) e^{-\kpa^2 \sigma_r^2}
\end{equation}
where $P_{\rm obs}$ was given in equation (\ref{eq:P1}).
In other words, the power is strongly suppressed for large $\kpa$.
The positional uncertainty is related to the redshift uncertainty $\sigma_z$
by $\sigma_r= c\sigma_z/\hz$.

The introduction of this suppression enters the Fisher matrix calculation
through its effect on the effective volume.
Modes with a relatively large $\kpa$ will be swamped by shot noise
and therefore give no leverage on the power spectrum measurements.
Only modes with $\kpa\sigma_r\lesssim1$ are useful.
Much lower shot noise allows one to retain modes of larger $\kpa$,
but one is fighting a Gaussian suppression.

Because the measurement of $\hz$ depends on modes aligned near the line
of sight, 
the suppressed contribution from modes with large $\kpa$ 
increases the error on $\hz$ significantly. 
The measurement of $\DA$ arises from more transverse modes, and modes
with $\kpa=0$ always exist to give some measurement of $\DA$.
However, for large $k$, only a thin slab of modes with 
$\kpa\lesssim 1/\sigma_r$ remain useful.  As the number of modes
will scale as $\sigma_r^{-1}$, we expect that the errors on $\DA$
will scale as $\sqrt{\sigma_r}$ for $\kmax \sigma_r \gg 1$. 

\begin{figure}[t]
\plotone{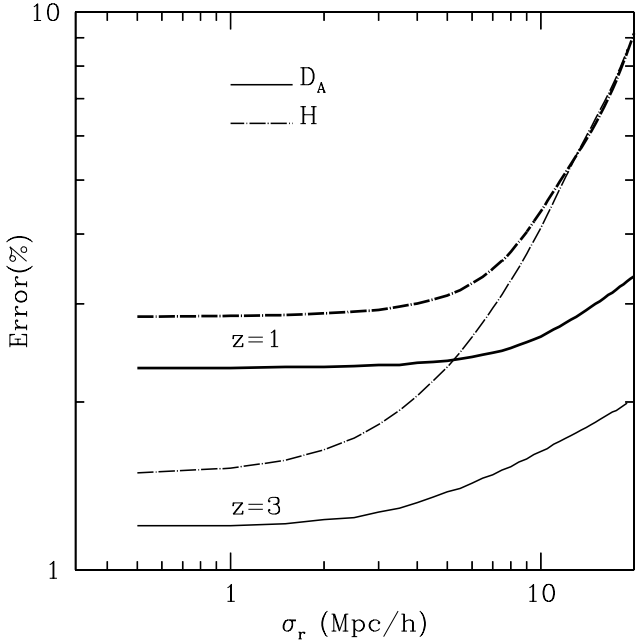}
\caption{\label{fig:sigr}
The error on $\DA$ and $\hz$ 
as a function of the line-of-sight positional uncertainty 
$\sigma_r$  for $z=1$ and $z=3$ redshift bins. CMB
and SDSS data are included in each redshift bin. $1\%$ of $\sigma_z/(1+z)$
corresponds to 34$h^{-1}{\Mpc}$ at $z=1$ and 27$h^{-1}{\Mpc}$ at $z=3$.}
\end{figure}

Figure \ref{fig:sigr} shows the fractional errors on $\hz$
and $\DA$ as a function of $\sigma_r$ for redshift surveys at
$z=1.0$ and $z=3$. 
The errors are constant for small $\sigma_r$
and then increase rapidly beyond a characteristic threshold. Performance
at $z=3$ degrades at smaller $\sigma_r$ than performance at $z=1.0$. 
This is because of the larger value of $\kmax$ for $z=3$. 
An additional small but non-zero effect is that the redshift
distortions are smaller at $z=3$ than at $z=1$.  Larger distortions
increase the power in the radial direction and allow modes with
slightly larger $\kpa$ to survive the shot noise.
The errors on $\hz$ degrade sharply for $\sigma_r\gtrsim10\hMpc$
at $z=1$ and $5\hMpc$ at $z=3$.  These correspond to redshift errors
$\sigma_z$ of 0.006 and 0.007, respectively.  In terms of 
wavelength resolution $\sigma_\lambda/\lambda$, these are 0.003 and
0.002.  Hence, our general result is that fractional errors of 
0.25\% in $1+z$ are required to recover $H(z)$.

In Figure \ref{fig:sigr}, 
the errors on $\DA$ in both redshift bins increase relatively
slowly at $\sigma_r\gtrsim 10\hMpc$
and achieve the predicted $\sqrt{\sigma_r}$ dependence at 
large $\sigma_r$.
Therefore, to calculate $\sigma_{D_A}$ with
$\sigma_r$ bigger than the values appeared in Figure \ref{fig:sigr},
we can use $\sqrt{\sigma_r}$ dependence to interpolate (up to the
limit of $\sigma_z/(1+z)\approx 4\%$). 
For numerical reasons, we assume redshift error of $1\%$ in photometric
redshift. This is too optimistic for a normal photometric redshift, but
one can scale to larger uncertainties. For example, a $4\%$ uncertainty
would have errors twice as large, which could be compensated by making
the survey area 4 times as large. $1\%$ errors in $(1+z)$ correspond to
$\sigma_r=34\hMpc$ at $z=1$ and $\sigma_r=27\hMpc$ at $z=3$.

\begin{figure*}[t]
\epsscale{2}
\plottwo{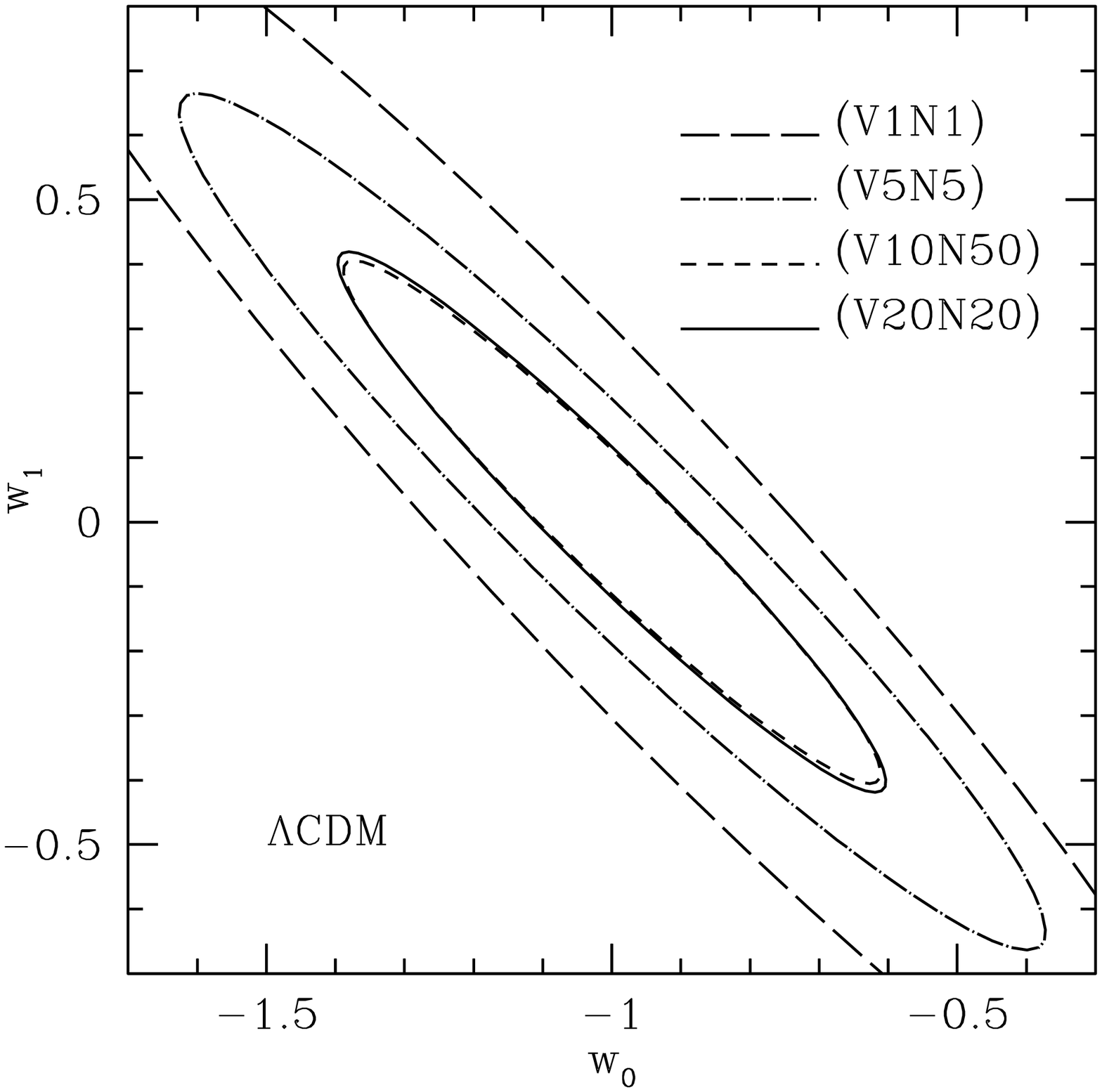}{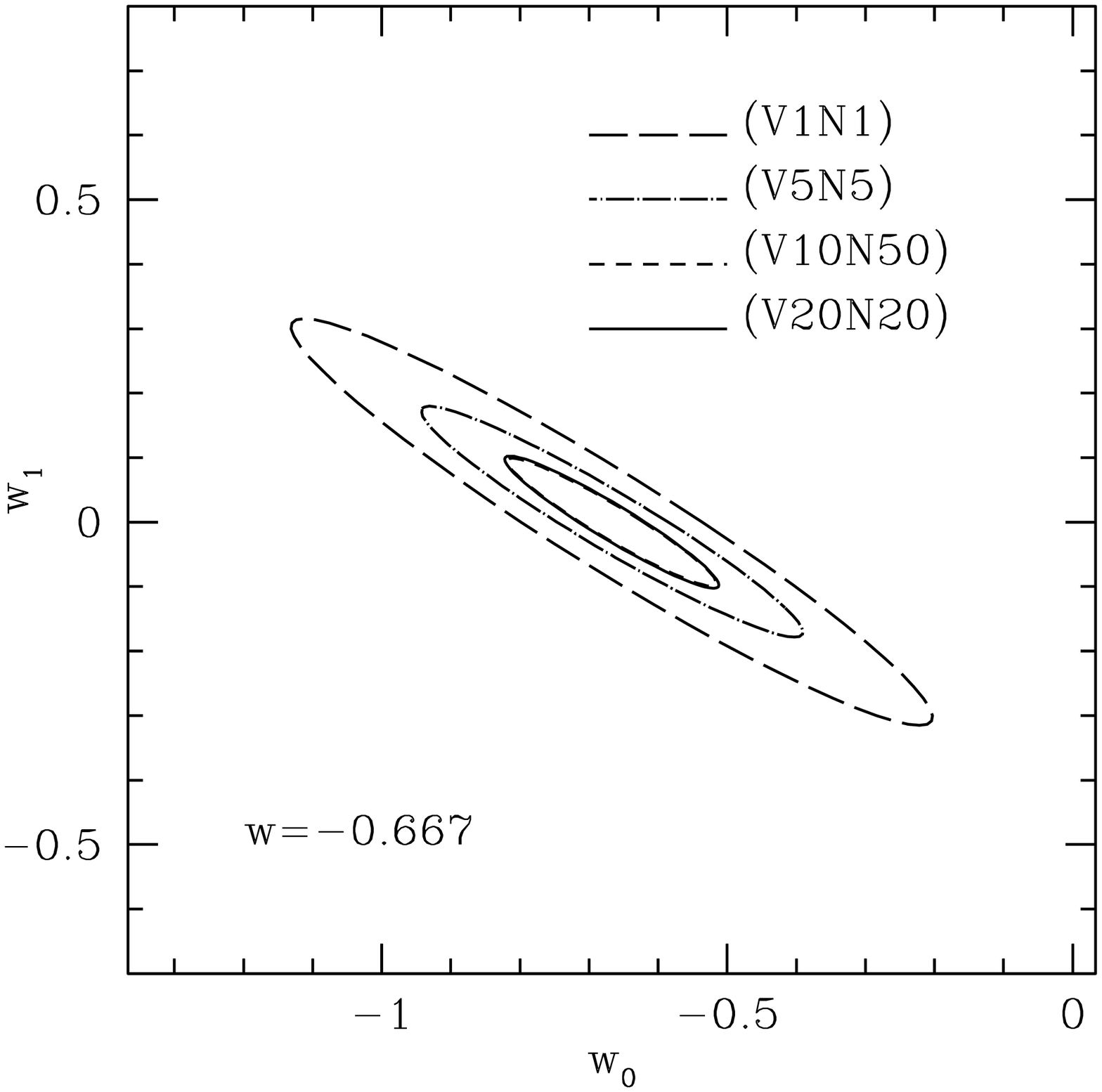}
\epsscale{1}
\caption{\label{fig:daep}
Elliptical error regions for $w_0$ and $w_1$ 
for photometric redshift surveys with $1\%$ of
$\sigma_z/(1+z)$ ($\sigma_z=0.02$ for $z\sim1$, $\sigma_z=0.04$ for
$z=3.0$).  Left panel: $\LCDM$.  Right panel: Model 2 ($w=-2/3$). 
CMB data and all redshift surveys are included in all cases. The
survey parameters written inside the parenthesis are for both $z\sim1$
and $z=3$ bins. V numbers specify the change in the survey volume 
relative to the baseline;
N the change in the number of galaxies. For SDSS, the fiducial survey
parameters (V1N1) with spectroscopic redshifts are used. In both figures, V10N50 is nearly the same as V20N20. }
\end{figure*}

Table \ref{tab:daonly1} shows the errors on $\DA$ and $\hz$ for
different survey conditions. Increasing the survey volume five-fold
while keeping the target density fixed (i.e. V5N5) decreases the 
error by $\sim \sqrt{5}$, as before. 
Further increases of  the survey volume, which are omitted
in Table \ref{tab:daonly1}, continue to follow the simple trend of
$\sqrt{\Vsur}$. Increasing the number density with fixed $\Vsur$ is
slightly more efficient than the spectroscopic redshift case ($\sigma_r
=0$) because the exponential suppression of modes with non-zero $\kpa$
means that there are always modes that benefit from a larger $n$ 
to achieve $nP \sim 1$.

Table \ref{tab:daonlyw} shows the propagated errors on dark energy
parameters for $\LCDM$. The left panel of Figure \ref{fig:daep} shows
the corresponding error ellipses. The errors on $\w_0$ and $\w_1$ using
photometric redshifts increases by a factor of $\sim2.4$ with the fiducial
condition (V1N1) relative to spectroscopic results while $\w(z_{\rm pivot})$
is less affected. As shown in Figure \ref{fig:daep}, this is because
the ellipse is more elongated with a relatively little increase in its
minor axis compared to left panel of Figure \ref{fig:ellip}. This is due
to the increased dominance of $z\sim1$ survey over $z=3$. $z_{\rm pivot}$
increases slightly with photometric redshift data. As shown in Figure
\ref{fig:deriv}, $\DA$ at higher $z$ contributes more to the information
compared to $\hz$. Thus, eliminating $\hz$ by using photometric redshifts
will weight higher redshifts slightly more.

Table \ref{tab:daonlyw} also shows that increasing $\Vsur$ by a factor of 20
(V20N20) or a factor of 10 with a 5-fold increase in target density $n$ (V10N50)
allows the results from photometric redshift surveys to achieve the
accuracy of the spectroscopic redshift surveys. 
This corresponds to 10 or 20 thousand square degrees of sky at $z=1$ 
(with 1\% errors in $1+z$). 
Observationally, increasing the
number density by 5 times (V10N50) will be more difficult than doubling
the survey area because the galaxy luminosity function flattens out
around this number density, so that $\sqrt{2}$ in depth is far less
than a factor of 5 in source counts. The errors of $\w_0$ and $\w_1$
do not scale trivially with $\sqrt{\Vsur}$ because the SDSS and CMB survey
parameters are being held fixed.

Results including the SNe survey data are shown in Table
\ref{tab:daonlywsnap}. 
With SNe, improving the redshift survey condition
to between V5N5 and V10N10 allows the photometric redshift survey
to recover the spectroscopic result in Table \ref{tab:resw}. This is
equivalent to an imaging survey of about 30,000 square degrees 
with $4\%$ photometric
redshift error at $z\sim1$ and a depth to reach 1000 $z\sim1$ galaxies
per square degree.  Surveys such as Pan-STARRS (http://pan-starrs.ifa.hawaii.edu) or the Large Synoptic Survey Telescope (http://www.lssto.org) could
achieve this.

Table \ref{tab:daonlywM2} and Table \ref{tab:daonlywsnapM2} show the same
analysis as Table \ref{tab:daonlyw} and \ref{tab:daonlywsnap}, but for
Model 2 ($\w=-2/3$) instead of $\LCDM$.  The degradation of performance
relative to the spectroscopic case is similar.
Like the $\LCDM$ case, V10N50 or V20N20 recovers the spectroscopic
result without SNe, and V5N5 or V10N10 does the job with SNe. 
The right panel of Figure \ref{fig:daep} shows the corresponding error ellipses.

Therefore, after considering both the projection and power suppression 
effects of redshift uncertainties, we expect that,
when combined with supernovae data, 
surveys with 4\% errors on $1+z$ and roughly 30 times more volume than our 
baseline surveys will be equivalent to the spectroscopic surveys.
As this is essentially the full sky at $z\sim1$, improving beyond
these levels will require better redshift accuracy.

\section{Conclusion}

Understanding the acceleration of the Universe is one of the most 
important problems in both cosmology and fundamental particle physics. 
Identifying the physical cause, whether dark energy or some alteration
to the theory of gravitation, is certain to be a major breakthrough.
Precision measurements of the expansion history of the universe could
be crucial in choosing between alternative theories. 
In this paper, we
demonstrated that a standard ruler test using baryonic acoustic
oscillations imprinted in the large scale structure could be 
a superb probe of the acceleration history.
The oscillations in the galaxy power spectrum are expected to 
be robust against contamination from clustering bias, redshift
distortions, and other broadband systematic errors.

We have studied the performance that could be achieved on dark energy
models from the measurement of the acoustic oscillations in large
galaxy spectroscopic surveys at redshifts 0.3, 1, and 3.
The $z\sim1$ baseline survey uses 900,000 galaxies to probe 
$1.7h^{-3}\Gpc^3$; the $z=3$ survey uses a half million galaxies
to cover $0.5h^{-3}\Gpc^3$.  While these numbers are large, the
number densities are not, which means that relatively bright galaxies
could be used.
Using a Fisher matrix treatment of the statistical errors 
that result from the three-dimensional power spectra, as well
as CMB and SNe data, we forecasted errors on the distances along
and across the line of sight and then projected these measurements
of $\hz$ and $\DA$ onto dark energy parameters.  Of course, the
cosmographical performance are independent of the details of the
dark energy model.  We summarize our major results below.

First, we have shown that (1--$\sigma$) errors of 0.037 on $\Ox$, 0.10
on $\w(z=0.8)$, and 0.28 on $dw/dz$ are achievable for $\LCDM$ when
CMB provides the scale of the baryonic oscillations. The constraints on
$dw/dz$ are comparable to those from the luminosity distances
of future SNe data.  Most of constraints were contributed by
information in the higher redshift surveys ($z \gtrsim 0.6$) because the
baryonic oscillations in the power spectrum are better preserved against
nonlinearity at higher redshift. When we combined the redshift survey
data with the SNe data, the constraints were improved
to 0.16 on $d\w/dz$.

Second, we found that fiducial dark energy models with less negative
$\w$ than $\LCDM$ improve overall performance and also favor the galaxy 
redshift surveys relative to the SNe data.
Together, a 0.05 measurement of $dw/dz$ is achieved!

Third, we discussed how the quality of constraints depends upon the 
the survey volume and number density. Increasing the survey
volume with the number density fixed always gives the better result by
$\sqrt{\Vsur}$. Increasing the number density, that is, going deeper with
the volume fixed, will also improve the constraints but with asymptotic
saturation.  Changing the survey volume with a fixed total number of objects
has a maximum in performance that is close to the baseline values.

Forth, we computed how well an imaging survey with photometric redshifts
could measure the acoustic oscillations.
We find that errors of 0.25\% in $1+z$ are necessary to retain 
information on the Hubble parameter $H(z)$.  However, redshift errors of 4\%
in $1+z$ can be tolerated without losing the oscillations to projection
effects, and the angular diameter distance could be measured as a function
of redshift.
We estimate that a survey 20 times larger than our baseline but
with 1\% redshift error on $1+z$ is needed to replace the spectroscopy,
but that the requirement drops to 5--10 times larger when combined
with the constraints from SNe.
4\% redshift errors require four times more volume. 

To date, much of the attention in cosmological probes of acceleration has 
rightly been given to the studies of distant supernovae.  The acoustic
oscillations in the galaxy power spectrum have not even been conclusively
detected yet.  Nevertheless, we are encouraged by
the result that the study of acoustic oscillations in large galaxy surveys
can achieve comparable performance to upcoming SNe data sets.  
Given the mystery and importance of the acceleration of the universe, 
it is crucial to have multiple experiments with independent
systematic errors.  Moreover, the ability to measure $H(z)$ directly and
to probe the expansion at higher redshifts ($z\approx 3$) opens the
possibility of detecting new surprises.  Although the cosmological
constant model is most easily probed at lower redshifts, given the woeful 
history of theoretical predictions for dark energy, it seems to us unwise to 
design experiments based too closely on the assumptions of $\LCDM$.

While the required redshift surveys are large, they are feasible within
the current decade.  8-meter ground based telescopes are sufficiently
sensitive, but currently lack the necessary highly multiplexed wide-field 
spectroscopic capability.  Instruments such as the KAOS concept
(http://\linebreak www.noao.edu/kaos) could perform these surveys in about a year
of observing.  The surveys would of course have many other science
applications, both for the study of galaxy evolution and for the 
search for more speculative features of the linear perturbations,
e.g.~primordial non-Gaussianity or additional preferred scales.
At $z=3$, the reach into the linear regime on intermediate scales
exceeds even that of the CMB.  Hence, we conclude that such surveys
are attractive options for the study of large-scale structure over 
the next decade.

\acknowledgements  
We thank Chris Blake, Arjun Dey, Karl Glazebrook,
Eric Linder, and Saul Perlmutter for useful discussions.
H.S. was supported by a University of Arizona College of Science Graduate Fellowship.
D.J.E. is supported by National Science Foundation grant AST-0098577
and by an Alfred P. Sloan Research Fellowship.

\clearpage
\onecolumn
\renewcommand{\textfraction}{0}

\begin{deluxetable}{l|l|cccccccc}
\tablewidth{0pt}
\tabletypesize{\footnotesize}
\tablecaption{\label{tab:con}Baseline Survey Parameters}
\startdata \hline\hline 
\tableskten
Survey& $z$ & $k_{\rm max}$ & $V_{\rm survey}\tablenotemark{a}$ & 
$N_{\rm gal}\tablenotemark{b}$ & bias$\tablenotemark{c}$ & 
$P(0.2h \iMpc)\tablenotemark{d}$ & $P(k_{\rm max})$ & $nP(0.2h\iMpc)$ 
& $nP(k_{\rm max})$\\ 
&& $(h\iMpc)$ & ($h^{-3} \Gpc^3)$ & $(10^5)$ && $h^{-3} \Mpc^3$ & 
	$h^{-3} \Mpc^3$ &&\\\tableskten\hline\tableskten
SDSS    & 0.3 & 0.11 & 1.0  & 1.0  & 2.13 &      & 22900 && 2.29 \\\tableskten\hline
\tableskten
$z\sim1$& 0.6 & 0.15 & 0.29 & 1.44 & 1.25 &      & 4660  && 2.33 \\\tableskten
        & 0.8 & 0.17 & 0.40 & 2.00 & 1.40 &      & 3590  && 1.80 \\\tableskten
        & 1.0 & 0.19 & 0.49 & 2.46 & 1.55 &      & 3090  && 1.55 \\\tableskten
        & 1.2 & 0.21 & 0.56 & 2.82 & 1.70 & 2860 & 2620  & 1.43 &1.31 \\\tableskten\hline\tableskten
$z=3$   & 3.0 & 0.53 & 0.50 & 5.0  & 3.30 & 2950 & 430   & 2.95&0.43 \\
\enddata
\tablenotetext{a}{1000 square degrees for $z \sim 1$; 
140 square degrees for $z=3$} 
\tablenotetext{b}{The number density $n$: 
$10^{-4}h^3\Mpc^{-3}$ for SDSS, 
$5\times10^{-4}h^3\Mpc^{-3}$ for $z\sim1$, 
and $10^{-3}h^3\Mpc^{-3}$ for $z=3$}
\tablenotetext{c}{Calculated using Equation (\protect\ref{eq:bias})
assuming $\sigma_{8,\rm mass}=0.9$ at $z=0$, $\sigma_{8,g}=1.8$ for SDSS, 
$\sigma_{8,g}=1$ for $z\sim1$ and $z=3$. }
\tablenotetext{d}{Powers at $k=0.2h\iMpc$ are slightly different for 
$z\sim1$ and $z=3$ because redshift distortions are
included in normalization of $\sigma_{8,g}$}
\end{deluxetable}

\begin{deluxetable}{lcc}
\tablewidth{0pt}
\tabletypesize{\small}
\tablecaption{\label{tab:model} Dark Energy Models}
\startdata \hline\hline \tableskth
Model & $w_0$ & $w_1$ \\ \tableskth\hline
\hspace*{1in}&\hspace*{1in}&\hspace*{1in}\\[-7pt] 
1 ($\Lambda$CDM) & $-1$   & 0$\tablenotemark{a}$ \\ \tableskth
2                & $-2/3$ & 0$\tablenotemark{a}$ \\ \tableskth
3                & $-2/3$ & $1/6\tablenotemark{b}$ \\ \tableskth
4                & $-1$   & $1/3\tablenotemark{b}$ \\ \tableskth
5                & $-4/3$ & $1/3\tablenotemark{b}$ \\ \tableskth
6                & $-1.15$ & $1/3\tablenotemark{b}$ \\
\enddata
\tablenotetext{a}{$w_1$ perturbations in these models were considered
to extend to $z=\infty$; 
however, the derivatives were computed with infinitesimal stepsizes,
so the $w>0$ region at high redshift was not an issue.}
\tablenotetext{b}{$w(z) = w_0 + w_1 z$ for $z<z_t$ and $w_0+w_1 z_t$
beyond.  We use $z_t=2$.}
\end{deluxetable}

\begin{deluxetable}{l|ccccccc}
\tablewidth{0pt}
\tabletypesize{\small}
\tablecaption{\label{tab:dahz}Marginalized Errors on $\DA$ and $\hz$ for $\LCDM$}
\startdata\hline\hline \tableskei
Redshift&0.3&0.6&0.8&1.0&1.2&3&1000\\ \tableskei\hline \tableskei
$\DA$ (\%) &5.19&4.30&3.22&2.30&2.03&1.19&0.219\\\tableskei
$\hz$ (\%) &5.80&5.19&3.59&2.84&2.53&1.48&\\
\enddata
\tablecomments{The fractional percentage errors (1--$\sigma$) on cosmological
distances from the combination of CMB, SDSS, and our standard surveys
at $z\sim1$ and $z=3$. }
\end{deluxetable}
\clearpage
\begin{deluxetable}{l|l||c|cc|cc|cc|cc|cc|cc}
\tablewidth{0pt}
\tabletypesize{\small}
\rotate
\tablecaption{\label{tab:cor}Correlation Matrix for Distance Measurements in $\LCDM$}
\startdata \hline \tableskft
\multicolumn{1}{l|}{}&\multicolumn{1}{l||}{z}& \multicolumn{1}{c}{\small{CMB}}&\multicolumn{2}{|c}{0.3}&\multicolumn{2}{|c}{0.6}& \multicolumn{2}{|c}{0.8}& \multicolumn{2}{|c}{1.0}& \multicolumn{2}{|c}{1.2}  & \multicolumn{2}{|c}{3.0}\\\tableskft \hline \tableskft
z& &$\DAA$&$\DAA$&$hzz$ & $\DAA$  &$\hzz$ & $\DAA$&  $\hzz$ & $\DAA$ & $\hzz$&$\DAA$&  $\hzz$ & $\DAA$ & $\hzz$ \\ \tableskft\hline\hline\tableskft
& $\sqrt{D_{ii}}$\tablenotemark{a}
& 0.002
& 0.052
& 0.058
& 0.043
& 0.052
& 0.032
& 0.036
& 0.023
& 0.028
& 0.020
& 0.025
& 0.012
& 0.015\\\tableskft \hline\hline\tableskft
{\small CMB}& $\DAA$
& 1.000
& 0.040
& $-$0.048
& 0.054
& $-$0.045
& 0.061
& $-$0.064
& 0.103
& $-$0.087
& 0.123
& $-$0.100
& 0.233
& $-$0.192
\\\tableskft \hline\tableskft
0.3& $\DAA$
& 0.040
& 1.000
& $-$0.256
& 0.002
& $-$0.002
& 0.003
& $-$0.003
& 0.005
& $-$0.004
& 0.006
& $-$0.005
& 0.011
& $-$0.009
\\\tableskft
& $\hzz$
& $-$0.048
& $-$0.256
& 1.000
& $-$0.003
& 0.003
& $-$0.003
& 0.004
& $-$0.006
& 0.005
& $-$0.007
& 0.006
& $-$0.013
& 0.011
\\ \tableskft\hline\tableskft
0.6& $\DAA$
& 0.054
& 0.002
& $-$0.003
& 1.000
& $-$0.255
& 0.004
& $-$0.004
& 0.006
& $-$0.005
& 0.008
& $-$0.006
& 0.015
& $-$0.012
\\\tableskft
& $\hzz$
& $-$0.045
& $-$0.002
& 0.003
& $-$0.255
& 1.000
& $-$0.003
& 0.003
& $-$0.005
& 0.005
& $-$0.006
& 0.005
& $-$0.012
& 0.010
\\\tableskft \hline\tableskft
0.8& $\DAA$
& 0.061
& 0.003
& $-$0.003
& 0.004
& $-$0.003
& 1.000
& $-$0.304
& 0.007
& $-$0.006
& 0.009
& $-$0.007
& 0.016
& $-$0.013
\\\tableskft
& $\hzz$
& $-$0.064
& $-$0.003
& 0.004
& $-$0.004
& 0.003
& $-$0.304
& 1.000
& $-$0.008
& 0.006
& $-$0.009
& 0.007
& $-$0.017
& 0.014
\\\tableskft \hline\tableskft
1.0& $\DAA$
& 0.103
& 0.005
& $-$0.006
& 0.006
& $-$0.005
& 0.007
& $-$0.008
& 1.000
& $-$0.124
& 0.015
& $-$0.012
& 0.028
& $-$0.023
\\\tableskft
& $\hzz$
& $-$0.087
& $-$0.004
& 0.005
& $-$0.005
& 0.005
& $-$0.006
& 0.006
& $-$0.124
& 1.000
& $-$0.012
& 0.010
& $-$0.024
& 0.020
\\ \tableskft\hline\tableskft
1.2& $\DAA$
& 0.123
& 0.006
& $-$0.007
& 0.008
& $-$0.006
& 0.009
& $-$0.009
& 0.015
& $-$0.012
& 1.000
& $-$0.120
& 0.033
& $-$0.028
\\\tableskft
& $\hzz$
& $-$0.100
& $-$0.005
& 0.006
& $-$0.006
& 0.005
& $-$0.007
& 0.007
& $-$0.012
& 0.010
& $-$0.120
& 1.000
& $-$0.027
& 0.023
\\ \tableskft\hline\tableskft
3.0& $\DAA$
& 0.233
& 0.011
& $-$0.013
& 0.015
& $-$0.012
& 0.016
& $-$0.017
& 0.028
& $-$0.024
& 0.033
& $-$0.027
& 1.000
& $-$0.203
\\\tableskft
& $\hzz$
& $-$0.192
& $-$0.009
& 0.011
& $-$0.012
& 0.010
& $-$0.013
& 0.014
& $-$0.023
& 0.020
& $-$0.028
& 0.023
& $-$0.203
& 1.000
\\ 

\enddata
\tablecomments{All terms are normalized by diagonal terms given in the 
first row: $a^\prime_{ij}=\frac{a_{ij}}{\sqrt{a_{ii}a_{jj}}}$}
\tablenotetext{a}{The square root of the diagonal terms of the covariance matrix.
These are the 1--$\sigma$ fractional percentage errors on these quantities.}
\end{deluxetable}

\begin{deluxetable}{lll|cccccc}
\tablewidth{0pt}
\tabletypesize{\small}
\tablecaption{\label{tab:resw}Marginalized Errors on Dark Energy Parameters for $\Lambda$CDM}
\startdata  \hline \hline \tableskni
$z\sim1$&$z=3$&SNe& $\sigma_{\Omega_m\;h^2}$/$\Omega_m\;h^2$&$\sigma_{\Omega_x}$& $\sigma_{w_0}$ & $\sigma_{w_1}$ & $z_{\rm pivot}$&$\sigma_{w_{zpivot}}$\\  \tableskni \hline  \tableskni
&& &0.0094&0.0926&0.882&1.172&0.729&0.218 \\ \tableskni 
$\surd$&& &0.0090&0.0378&0.281&0.353&0.735&0.107 \\ \tableskni 
&$\surd$& &0.0086&0.0758&0.466&0.446&0.959&0.184 \\\tableskni 
$\surd$&$\surd$& &0.0083&0.0368&0.245&0.280&0.796&0.102 \\ \tableskni 
\hline \tableskni
&&$\surd$ &0.0093&0.0088&0.116&0.231&0.478&0.035 \\ \tableskni 
$\surd$&&$\surd$ &0.0088&0.0083&0.093&0.183&0.471&0.033 \\ \tableskni 
&$\surd$&$\surd$ &0.0086&0.0085&0.096&0.189&0.479&0.034 \\ \tableskni 
$\surd$&$\surd$&$\surd$ &0.0082&0.0082&0.083&0.161&0.476&0.032 \\ 
\enddata
\tablecomments{Check marks indicate the data sets being used; 
CMB and SDSS data are included in all sets. 
The fiducial redshift survey parameters (V1N1) are used.
$z_{\rm pivot}$ is the redshift at which the errors on the 
value of $w(z)$ is independent from the slope $w_1$.
$\sigma(w_{zpivot})$ is the error on the value of $w$ at that redshift;
this is also the error on $w$ that would be found if $w_1$ were held
fixed at the fiducial value.
All errors are 1--$\sigma$.
}
\end{deluxetable}

\clearpage
\renewcommand{\arraystretch}{1.11}
\begin{deluxetable}{l|lll|cccccc}
\tablewidth{0pt}
\tabletypesize{\small}
\tablecaption{\label{tab:w}Cosmological Errors from Different Fiducial Dark Energy Models}
\startdata  \hline \hline \tableskte
Model  &$z\sim1$&$z=3$&SNe& $\sigma_{\Omega_m\;h^2}$/$\Omega_m\;h^2$&$\sigma_{\Omega_x}$& $\sigma_{w_0}$ & $\sigma_{w_1}$ & $z_{\rm pivot}$&$\sigma_{w_{zpivot}}$ \\ \tableskte\hline\tableskte
Model 2
&$\surd$&& &0.0090&0.0259&0.129&0.092&1.249&0.057 \\\relax
&&$\surd$& &0.0088&0.0485&0.247&0.158&1.484&0.078 \\\relax
&$\surd$&$\surd$& &0.0085&0.0245&0.119&0.081&1.333&0.050 \\
\hline\tableskte
&&&$\surd$ &0.0094&0.0134&0.087&0.122&0.682&0.023 \\\relax
&$\surd$&&$\surd$ &0.0090&0.0088&0.049&0.062&0.714&0.022 \\\relax
&&$\surd$&$\surd$ &0.0087&0.0097&0.056&0.072&0.713&0.023 \\\relax
&$\surd$&$\surd$&$\surd$ &0.0084&0.0083&0.045&0.052&0.744&0.022 \\
\hline\tableskte
Model 3
&$\surd$&& &0.0090&0.0266&0.131&0.097&1.306&0.033 \\\relax
&&$\surd$& &0.0088&0.0530&0.268&0.190&1.389&0.043 \\\relax
&$\surd$&$\surd$& &0.0085&0.0256&0.124&0.089&1.351&0.028 \\
\hline\tableskte
&&&$\surd$ &0.0094&0.0158&0.081&0.107&0.741&0.019 \\\relax
&$\surd$&&$\surd$ &0.0090&0.0092&0.045&0.049&0.853&0.017 \\\relax
&&$\surd$&$\surd$ &0.0087&0.0103&0.051&0.058&0.821&0.018 \\\relax\relax
&$\surd$&$\surd$&$\surd$ &0.0085&0.0086&0.042&0.042&0.906&0.016 \\
\hline\tableskte
Model 4
&$\surd$&& &0.0090&0.0320&0.180&0.144&1.203&0.049 \\\relax
&&$\surd$& &0.0088&0.0653&0.367&0.275&1.314&0.065 \\\relax
&$\surd$&$\surd$& &0.0085&0.0313&0.171&0.132&1.259&0.043 \\
\hline\tableskte
&&&$\surd$ &0.0094&0.0127&0.106&0.142&0.729&0.022 \\\relax
&$\surd$&&$\surd$ &0.0090&0.0090&0.062&0.075&0.774&0.021 \\\relax
&&$\surd$&$\surd$ &0.0087&0.0098&0.070&0.087&0.764&0.022 \\\relax
&$\surd$&$\surd$&$\surd$ &0.0085&0.0086&0.056&0.064&0.803&0.021 \\
\hline\tableskte
Model 5
&$\surd$&& &0.0090&0.0454&0.353&0.429&0.773&0.122 \\\relax
&&$\surd$& &0.0086&0.0844&0.560&0.547&0.968&0.183 \\\relax
&$\surd$&$\surd$& &0.0083&0.0443&0.312&0.344&0.846&0.113 \\
\hline\tableskte
&&&$\surd$ &0.0094&0.0089&0.132&0.258&0.488&0.040 \\\relax
&$\surd$&&$\surd$ &0.0089&0.0085&0.106&0.206&0.480&0.038 \\\relax
&&$\surd$&$\surd$ &0.0086&0.0087&0.109&0.208&0.491&0.038 \\\relax
&$\surd$&$\surd$&$\surd$ &0.0083&0.0084&0.095&0.179&0.487&0.037 \\
\hline\tableskte
Model 6
&$\surd$&& &0.0090&0.0368&0.240&0.232&0.975&0.078 \\\relax
&&$\surd$& &0.0087&0.0766&0.463&0.386&1.166&0.107 \\\relax
&$\surd$&$\surd$& &0.0084&0.0365&0.224&0.201&1.057&0.070 \\
\hline\tableskte
&&&$\surd$ &0.0094&0.0101&0.119&0.190&0.604&0.029 \\\relax
&$\surd$&&$\surd$ &0.0090&0.0086&0.081&0.125&0.609&0.029 \\\relax
&&$\surd$&$\surd$ &0.0087&0.0090&0.087&0.133&0.619&0.028 \\\relax
&$\surd$&$\surd$&$\surd$ &0.0084&0.0084&0.071&0.105&0.626&0.028 
\enddata
\tablecomments{Check marks indicate the data sets being used; 
CMB and SDSS data are included in all sets. 
The fiducial redshift survey parameters (V1N1) are used.}
\end{deluxetable}
\renewcommand{\arraystretch}{1.0}

\clearpage
\begin{deluxetable}{ll|c|cccccc}
\tablewidth{0pt}
\tabletypesize{\small}
\tablecaption{\label{tab:dahz1}Marginalized Errors on $\DA$ and $\hz$ as a Function of Survey Parameters }
\startdata \hline\hline \tableskni
\multicolumn{2}{c}{Surveys}& \multicolumn{1}{|c}{}& \multicolumn{6}{|c}{Redshift}  \\  \tableskni\hline  \tableskni
$z\sim1$&$z=3$&&0.3&0.6&0.8&1.0&1.2&3.0\\\tableskni\hline\tableskni
 V1N1 & V1N1& $\DA$&5.19&4.30&3.22&2.30&2.03&1.19 \\\tableskni
&& $\hz$&5.80&5.19&3.59&2.84&2.53&1.48 \\  \tableskni\hline \tableskni

 V1N5 & V1N5& $\DA$&5.19&3.50&2.57&1.74&1.52&0.88 \\\tableskni
&& $\hz$&5.80&4.44&3.00&2.30&2.01&1.08 \\ \tableskni \hline \tableskni

 V5N1 & V5N1& $\DA$&5.19&3.52&2.75&2.12&1.91&1.10 \\\tableskni
&& $\hz$&5.80&3.83&2.79&2.32&2.13&1.33 \\  \tableskni\hline \tableskni

 V5N5 &V5N5& $\DA$&5.19&1.93&1.45&1.04&0.93&0.57 \\\tableskni
&& $\hz$&5.80&2.33&1.62&1.28&1.15&0.69 \\ 
\enddata
\tablecomments{1--$\sigma$ fractional percentage errors on cosmological
distances.
CMB and SDSS are included in all sets. 
For SDSS, the fiducial condition (V1N1) is always used. 
V1N1 means the fiducial condition described in Table \ref{tab:con}. 
V5N1: 5 times larger survey volume with 5 times smaller number density. 
V1N5: 5 times higher number of objects with the standard survey volume, 
i.e. 5 times higher number density. 
V5N5: 5 times more survey volume with the standard number density.}
\end{deluxetable}

\begin{deluxetable}{ll|cccccc}
\tablewidth{0pt}
\tabletypesize{\small}
\tablecaption{\label{tab:surveycon}Marginalized errors for $\Lambda$CDM for Various Survey Sizes}
\startdata \hline\hline \tableskei
$z\sim1$&$z=3$& $\sigma_{\Omega_m\;h^2}$/$\Omega_m\;h^2$&$\sigma_{\Omega_x}$& $\sigma_{w_0}$ & $\sigma_{w_1}$ & $z_{\rm pivot}$&$\sigma_{w_{zpivot}}$\\  \tableskei\hline \tableskei
 V1N1&      &0.0090&0.0378&0.281&0.353&0.735&0.107 \\\tableskei

 V5N5&      &0.0079&0.0195&0.142&0.191&0.680&0.056 \\\tableskei
     & V1N1 &0.0086&0.0758&0.466&0.446&0.959&0.184 \\\tableskei

     & V5N5 &0.0070&0.0724&0.399&0.306&1.239&0.126 \\ \tableskei\hline \tableskei
 V1N1& V1N1 &0.0083&0.0368&0.245&0.280&0.796&0.102 \\\tableskei

 V1N1& V5N5 &0.0069&0.0358&0.210&0.199&0.961&0.088 \\\tableskei

 V5N5& V1N1 &0.0074&0.0192&0.135&0.176&0.699&0.055 \\\tableskei
 V5N5& V5N5 &0.0064&0.0186&0.120&0.142&0.762&0.053 \\
\enddata
\tablecomments{Left two columns indicate how the sizes of the $z\sim1$ and
$z=3$ surveys are being varied; blanks mean that the survey is excluded.
CMB and SDSS are included in all rows. 
For SDSS, the fiducial parameters (V1N1) are always used.}
\end{deluxetable}

\begin{deluxetable}{l|l|cccccc}
\tablewidth{0pt}
\tabletypesize{\small}
\tablecaption{\label{tab:b0.005}Marginalized Errors on $\DA$ and $\hz$ with Negligible $\beta$ for Different Baryon Fractions}
\startdata\hline\hline \tableskei
\multicolumn{1}{l}{$\Omega_b$}& \multicolumn{1}{|l}{}& \multicolumn{6}{|c}{Redshift}  \\  \tableskei\hline \tableskei
&&0.3&0.6&0.8&1.0&1.2&3.0\\ \tableskei \hline \tableskei
0.005&$\DA$&13.03&15.64&11.47&9.21&7.68&3.16\\\tableskei
&$\hz$&12.71&14.05&10.50&8.49&7.27&3.56\\ \tableskei\hline \tableskei
0.05&$\DA$&5.80&5.66&4.03&2.92&2.59&1.44 \\\tableskei
&$\hz$&6.68&7.83&4.75&4.04&3.64&2.03\\
\enddata
\tablecomments{1--$\sigma$ fractional percentage errors on cosmological
distances for $\LCDM$. CMB and SDSS are included in all sets. The derivatives are computed with $\beta\approx 0$, thereby causing the redshift and cosmological distortions to be more degenerate. The usual $\beta$'s are used to compute $\Veff$.}
\end{deluxetable}

\begin{deluxetable}{ll|c|cccccc}
\tablewidth{0pt}
\tabletypesize{\small}
\tablecaption{\label{tab:daonly1}Marginalized Errors on $\DA$ and $\hz$ 
for Photometric Redshift Surveys}
\startdata \hline\hline \tableskni
\multicolumn{2}{l}{Surveys}& \multicolumn{1}{|c}{}& \multicolumn{6}{|c}{Redshift}  \\ \tableskni \hline \tableskni
$z\sim1$ & $z=3$ & &0.3&0.6&0.8&1.0&1.2&3.0\\\tableskni\hline\tableskni
V1N1 &V1N1& $\DA$&5.19&6.98&4.84&4.25&3.90&2.26 \\ \tableskni
&& $\hz$&5.80&22.37&19.58&18.28&17.77&16.15 \\ \tableskni\hline\tableskni

V1N5 &V1N5& $\DA$&5.19&4.84&3.26&2.81&2.56&1.57 \\ \tableskni
&& $\hz$&5.80&16.33&14.29&13.27&12.88&10.11 \\ \tableskni\hline\tableskni

V5N5 &V5N5& $\DA$&5.19&3.14&2.19&1.93&1.77&1.06 \\ \tableskni
&& $\hz$&5.80&10.02&8.78&8.21&7.99&7.26 \\  

\enddata
\tablecomments{The fractional percentage error on the cosmological distances
under conditions appropriate to photometric redshifts.
The redshift accuracy has been degraded to 1\% (1--$\sigma$) on 
$\sigma_z/(1+z) = \Delta\lambda/\lambda$, 
i.e. $\sigma_z = 2\%$ at $z=1$ and 4\% at $z=3$.
The results will scale as $\sqrt{\sigma_z}$, and $\sigma_z$ would
typically be larger for actual photometric redshifts.
The left two columns show variations in the survey parameters.
V1N1 is the standard survey volume and number density.
V1N5 allows for a 5-fold increase in the number density.
V5N5 is 5 times more volume at the standard number density. 
CMB and SDSS are included in all sets. 
For SDSS, the fiducial survey parameters (V1N1) are used and spectroscopic
redshifts ($\sigma_z=0$) are adopted. 
The $\Lambda$CDM fiducial model is used. }
\end{deluxetable}

\begin{deluxetable}{ll|cccccc}
\tablewidth{0pt}
\tabletypesize{\small}
\tablecaption{\label{tab:daonlyw}Marginalized Errors for $\Lambda$CDM 
for Photometric Redshift Surveys}
\startdata \hline \hline\tableskei
$z\sim1$& $z=3$ & $\sigma_{\Omega_m\;h^2}$/$\Omega_m\;h^2$&$\sigma_{\Omega_x}$ & $\sigma_{w_0}$ & $\sigma_{w_1}$ & $z_{\rm pivot}$&$\sigma_{w_{zpivot}}$\\ \tableskei\hline\tableskei
V1N1& &0.0091&0.0799&0.613&0.704&0.834&0.175 \\\tableskei
V1N5& &0.0089&0.0736&0.534&0.578&0.887&0.150 \\\tableskei

&V1N1 &0.0090&0.0859&0.752&0.979&0.735&0.218 \\\tableskei
&V1N5 &0.0089&0.0820&0.670&0.855&0.742&0.217 \\\tableskei
\hline\tableskei
V1N1&V1N1 &0.0088&0.0777&0.584&0.668&0.834&0.175 \\\tableskei

V1N5&V1N5 &0.0085&0.0708&0.504&0.544&0.887&0.148 \\\tableskei

V5N5&V5N5 &0.0077&0.0592&0.413&0.437&0.904&0.118 \\\tableskei
V10N10&V10N10 &0.0069&0.0485&0.335&0.352&0.912&0.094 \\\tableskei
V10N50&V10N50 &0.0062&0.0375&0.256&0.267&0.921&0.071 \\\tableskei
V20N20&V20N20 &0.0060&0.0379&0.261&0.276&0.908&0.073 \\
\enddata
\tablecomments{Redshift uncertainties have been applied as in 
Table \protect{\ref{tab:daonly1}}: 1\% in $\sigma_z/(1+z)$.
Various survey sizes are investigated, as detailed in the left two
columns.  V numbers specify the change in the survey volume; 
N the change in the number of galaxies.
Note that V20 is 20,000 square degrees at $z\sim1$ and about 3000
square degrees at $z\sim3$.
Blanks indicate that a survey has been excluded.
Larger redshift uncertainties can be offset with more volume;
for example, 4\% errors in $\sigma_z/(1+z)$ would require 4 times
more volume to produce the results in this Table.
CMB and SDSS are included in all rows. 
For SDSS, the fiducial parameters (V1N1) with spectroscopic redshifts
are used in all cases. }
\end{deluxetable}

\begin{deluxetable}{ll|cccccc}
\tablewidth{0pt}
\tabletypesize{\small}
\tablecaption{\label{tab:daonlywsnap}Marginalized Errors for $\Lambda$CDM for Photometric Redshift Survey with SNe}
\startdata \hline \hline\tableskei
$z\sim1$& $z=3$& $\sigma_{\Omega_m\;h^2}$/$\Omega_m\;h^2$&$\sigma_{\Omega_x}$& $\sigma_{w_0}$ & $\sigma_{w_1}$ & $z_{\rm pivot}$&$\sigma_{w_{zpivot}}$\\ \tableskei \hline\tableskei
& &0.0093&0.0088&0.116&0.231&0.478&0.035 \\\tableskei\hline\tableskei
V1N1& &0.0089&0.0085&0.106&0.212&0.473&0.034 \\\tableskei
V1N5& &0.0087&0.0083&0.098&0.195&0.470&0.034 \\\tableskei

&V1N1 &0.0090&0.0087&0.114&0.229&0.477&0.034 \\\tableskei
&V1N5 &0.0087&0.0086&0.113&0.227&0.476&0.034 \\\tableskei
\hline\tableskei
V1N1&V1N1 &0.0086&0.0084&0.105&0.211&0.473&0.034 \\\tableskei

V1N5&V1N5 &0.0083&0.0082&0.097&0.193&0.470&0.033 \\\tableskei

V5N5&V5N5 &0.0073&0.0080&0.087&0.171&0.471&0.032 \\\tableskei
V10N10&V10N10 &0.0065&0.0078&0.077&0.148&0.478&0.031 \\\tableskei
V10N50&V10N50 &0.0060&0.0076&0.066&0.119&0.496&0.030 \\\tableskei
V20N20&V20N20 &0.0058&0.0076&0.068&0.124&0.495&0.030 \\
\enddata
\tablecomments{As Table \protect\ref{tab:daonlyw}, 
but the SNe data have also been included.}
\end{deluxetable}
 
\begin{deluxetable}{ll|cccccc}
\tablewidth{0pt}
\tabletypesize{\small}
\tablecaption{\label{tab:daonlywM2} Marginalized Errors for Model 2 for Photometric Redshift Surveys}
\startdata \hline\hline\tableskei
$z\sim1$& $z=3$ & $\sigma_{\Omega_m\;h^2}$/$\Omega_m\;h^2$&$\sigma_{\Omega_x}$& $\sigma_{w_0}$ & $\sigma_{w_1}$ & $z_{\rm pivot}$&$\sigma_{w_{zpivot}}$\\ \tableskei\hline\tableskei
V1N1& &0.0091&0.0637&0.395&0.278&1.383&0.087 \\\tableskei
V1N5& &0.0090&0.0563&0.332&0.224&1.452&0.069 \\\tableskei

&V1N1 &0.0092&0.0618&0.342&0.249&1.285&0.122 \\\tableskei
&V1N5 &0.0090&0.0590&0.284&0.187&1.376&0.119 \\\tableskei
\hline\tableskei
V1N1&V1N1 &0.0090&0.0545&0.305&0.207&1.412&0.086 \\\tableskei

V1N5&V1N5 &0.0087&0.0457&0.245&0.160&1.476&0.068 \\\tableskei

V5N5&V5N5 &0.0079&0.0346&0.182&0.118&1.476&0.053 \\\tableskei
V10N10&V10N10 &0.0071&0.0264&0.138&0.090&1.463&0.041 \\\tableskei
V10N50&V10N50 &0.0064&0.0191&0.099&0.065&1.453&0.030 \\\tableskei
V20N20&V20N20 &0.0063&0.0195&0.102&0.067&1.436&0.032 \\
\enddata
\tablecomments{As Table \protect\ref{tab:daonlyw}, 
but fiducial model 2 ($w=-2/3$) has been used. }
\end{deluxetable}

\begin{deluxetable}{ll|cccccc}
\tablewidth{0pt}
\tabletypesize{\small}
\tablecaption{\label{tab:daonlywsnapM2} Marginalized Errors for Model 2 for Photometric Redshift Surveys with SNe}
\startdata \hline \hline\tableskei
$z\sim1$& $z=3$ & $\sigma_{\Omega_m\;h^2}$/$\Omega_m\;h^2$&$\sigma_{\Omega_x}$& $\sigma_{w_0}$ & $\sigma_{w_1}$ & $z_{\rm pivot}$&$\sigma_{w_{zpivot}}$\\ \tableskei\hline\tableskei
& &0.0094&0.0134&0.087&0.122&0.682&0.023 \\\tableskei\hline\tableskei
V1N1& &0.0091&0.0104&0.066&0.089&0.695&0.023 \\\tableskei
V1N5& &0.0089&0.0091&0.057&0.072&0.716&0.023 \\\tableskei

&V1N1 &0.0091&0.0120&0.076&0.107&0.678&0.023 \\\tableskei
&V1N5 &0.0089&0.0112&0.068&0.095&0.676&0.023 \\\tableskei
\hline\tableskei                       
V1N1&V1N1 &0.0088&0.0100&0.062&0.084&0.693&0.023 \\\tableskei

V1N5&V1N5 &0.0086&0.0088&0.053&0.067&0.713&0.023 \\\tableskei

V5N5&V5N5 &0.0077&0.0079&0.046&0.054&0.743&0.022 \\\tableskei
V10N10&V10N10 &0.0070&0.0073&0.041&0.044&0.799&0.021 \\\tableskei
V10N50&V10N50 &0.0063&0.0068&0.037&0.035&0.904&0.019 \\\tableskei
V20N20&V20N20 &0.0062&0.0069&0.038&0.037&0.881&0.020 \\
\enddata
\tablecomments{As Table \protect\ref{tab:daonlyw}, but fiducial model 2 ($w=-2/3$) has been used and SNe data is included.}
\end{deluxetable}
\clearpage

\end{document}